\documentclass[
  reprint,
  superscriptaddress,
  amsmath,amssymb,
  pra,
  floatfix
]{revtex4-2}
\usepackage{kotex}
\usepackage{booktabs}
\usepackage[T1]{fontenc}
\usepackage{ulem}
\usepackage{graphicx}
\usepackage{xcolor}
\usepackage{dcolumn}
\usepackage{bm}
\usepackage{hyperref}
\usepackage{amsmath}
\usepackage{amsfonts}
\usepackage{braket}
\usepackage{tcolorbox}

\usepackage{array}
\usepackage{multirow}

\begin{document}

\title{Quantum Photonic Time Crystals:\\ From Temporal Boundaries to Floquet Light--Matter Interactions}
\author{Younsung Kim}
\thanks{These authors contributed equally to this work.}
\affiliation{Department of Physics, Korea Advanced Institute of Science and Technology, Daejeon 34141, Republic of Korea}

\author{Kyungmin Lee}
\thanks{These authors contributed equally to this work.}
\affiliation{Department of Physics, Korea Advanced Institute of Science and Technology, Daejeon 34141, Republic of Korea}

\author{Kun Woo Kim}
\email{kunx@cau.ac.kr}
\affiliation{Department of Physics, Chung-Ang University, 06974 Seoul, Republic of Korea}

\author{Bumki Min}
\email{bmin@kaist.ac.kr}
\affiliation{Department of Physics, Korea Advanced Institute of Science and Technology, Daejeon 34141, Republic of Korea}

\date{\today}

\begin{abstract}
Photonic time crystals (PTCs) are temporally periodic media whose Floquet spectra can exhibit momentum gaps, parametric amplification, and effective non-Hermitian descriptions, making them an idealized setting for vacuum amplification and nonequilibrium light--matter dynamics. Their classical electrodynamics is now well developed; the quantum side is less so, and this focused review is an attempt to organize what exists. We trace that account from temporal boundaries to homogeneous Floquet media and light--matter dynamics. A single temporal boundary induces Bogoliubov mode mixing and photon-pair creation; in homogeneous bulk media, momentum conservation isolates counter-propagating \((k,-k)\) sectors and yields a two-mode \(SU(1,1)\) squeezing structure. Temporal periodicity promotes this to a Floquet problem with band and momentum-gap regimes, compactly described in a fixed Nambu basis. We then relate PTCs to the dynamical Casimir effect and parametric amplification, which share the same pair-creation mechanism but organize it through discrete resonances rather than a momentum-resolved bulk spectrum. We close with light--matter settings: spontaneous-emission decay and modulation-assisted excitation, atom--PTC dynamics, LDOS-based observables and their limits, and finite, dispersive, and experimentally accessible platforms.

\end{abstract}

\maketitle


\section{Introduction}

Time-varying electromagnetic media extend our control of waves by breaking time-translation symmetry. When the medium remains spatially uniform, spatial translation symmetry survives and the wavevector stays a conserved label~\cite{morgenthaler1958velocity,fante1971transmission,felsen1970wave}. Recent symmetry and conservation-law analyses place this momentum conservation within the broader context of uniform time-varying and space--time-varying media~\cite{ortegaGomez2023momentumConservation,caloz2019spacetime,liberal2024spatiotemporalSymmetries}. The simplest such structure is a single temporal boundary. There, the same symmetry gives rise to temporal refraction and reflection, the temporal counterparts of scattering at a spatial interface: the wavevector is preserved while the frequency is converted~\cite{xiao2014reflection,plansinis2015temporal,lee2018linear,lee2022resonance,moussa2023observation,ptitcyn2025temporal}. Recent optical time-interface experiments have revealed richer effects, such as temporal diffraction and time-modulated nonlinear frequency conversion~\cite{tirole2023double,tirole2024second}.

Photonic time crystals (PTCs), historically called temporal photonic crystals, are a distinctive periodically driven member of this family. The study of periodically time-varying media dates back to Morgenthaler's work on time-dependent permittivity and permeability, followed by Floquet treatments of time--space-periodic and purely time-periodic media by Cassedy, Oliner, Holberg, and Kunz~\cite{morgenthaler1958velocity,cassedy1963dispersion,cassedy1967dispersion,holberg1966parametric}. By the late 2000s, Zurita-S\'anchez, Halevi, and Cervantes-Gonz\'alez had recast these ideas in the modern optical language of temporal photonic crystals, now more commonly called photonic time crystals~\cite{zurita2009reflection}. Throughout this focused review, ``photonic time crystal'' refers to a photonic medium with an externally prescribed temporal periodicity, to be distinguished from Wilczek-type many-body time crystals, whose central concern is the spontaneous breaking of time-translation symmetry~\cite{wilczek2012quantum,PhysRevLett.109.160402,PhysRevLett.117.090402, PhysRevLett.118.030401,Choi2017,Zhang2017,doi:10.1126/science.abo3382,kyung2026selforganizedphotonictimequasicrystal}.

At the classical level, modern studies of PTCs are typically framed within a momentum-resolved Floquet-quasifrequency picture, where temporal periodicity can give rise to momentum gaps, parametric amplification, exceptional-point physics, and pronounced gap-edge responses~\cite{lustig2018topological,PhysRevB.98.085142,Park_OL_spatiotemporal,galiffi2022photonics,lustig2023photonic,asgari2024theory}. Early work by the Halevi group developed the resonant-response and pulse-propagation behavior of time-periodic dielectric slabs~\cite{zurita2010resonances,zurita2012pulse}, and the dynamic transmission-line experiment of Reyes-Ayona and Halevi then provided experimental evidence for a genuine wave-vector gap in a temporal-photonic-crystal analog operating in the long-wavelength limit~\cite{reyes2015observation}.
Non-Hermitian and topological viewpoints place momentum gaps, exceptional points, and topological structure at the center of PTC models~\cite{lustig2018topological,PhysRevB.98.085142,park2022revealing}. Related work situates PTCs within the broader landscape of spatiotemporal photonic crystals, where temporal momentum gaps can coexist with spatial energy gaps, and clarifies how PTC amplification relates to conventional parametric oscillation and amplification~\cite{Park_OL_spatiotemporal,lee2021parametric,sharabi2022spatiotemporal,khurgin2024photonic}.

Progress toward experimentally accessible PTC physics has also come through metasurface implementations and resonance-assisted strategies for enlarging momentum gaps in bulk media and metasurfaces~\cite{park2022revealing,wang2023metasurface,wang2025expanding}. A dynamically modulated microwave transmission-line experiment has demonstrated \(k\)-gap amplification and a temporal topological edge state~\cite{xiong2025observation}.
Classical and semiclassical studies link PTC physics to emitter- and LDOS-related phenomena. Within a classical emitter-based Floquet framework, temporal modulation was shown to reshape spontaneous-emission decay and to reveal a modulation-assisted excitation channel. Alongside these theoretical results, recent microwave experiments have accessed LDOS-based spectra and lasing-like self-oscillation in PTC platforms~\cite{park2025spontaneous,lee2026analogs}. These developments establish PTCs as distinct Floquet media with their own band-structure language, instability windows, and experimentally accessible signatures, and they provide the classical backdrop against which the quantum perspective developed below is set.

The quantum problem starts when the field amplitudes in this Floquet problem are promoted to operators. In time-varying media, temporal modulation induces Bogoliubov mixing between the positive- and negative-frequency sectors of the field operators, which appears quantum mechanically as two-mode squeezing and photon-pair creation from the vacuum. Historically, one early route to quantum radiation from time-dependent electromagnetic environments came from moving-boundary and cavity-modulation problems. Starting with Moore's variable-length cavity and extended in the moving-mirror analyses of Fulling and Davies, this work established that nonadiabatic boundaries can generate radiation from the vacuum~\cite{moore1970quantum,fulling1976radiation,davies1977radiation}. These ideas later became central to the language of the dynamical Casimir effect (DCE), in which nonadiabatic modulation of an electromagnetic environment induces Bogoliubov mixing and photon creation~\cite{nation2012colloquium,dodonov2020fifty}.

A complementary material-based route was proposed by Yablonovitch in 1989: a rapidly varying refractive medium could mix positive- and negative-frequency components of the field and turn zero-point fluctuations into real photons~\cite{PhysRevLett.62.1742}. Direct quantization schemes for nonstationary electromagnetic media then showed, already in simple linear dielectric models, that time-dependent media can support quantum vacuum effects, photon creation, and squeezing without any mechanically moving boundary~\cite{dodonov_PhysRevA.47.4422,law1994effective,PhysRevA.53.1031,PhysRevA.55.62}. Structurally, this shared quadratic bosonic dynamics places time-varying media in close contact with parametric amplification and two-mode squeezing, making DCE language and the \(SU(1,1)\) structure natural reference points for the quantum theory of PTCs~\cite{caves1982quantum,caves1985new,schumaker1985new,yurke19862}.

The quantum wave-mixing physics of time-varying media is most cleanly exposed in the elementary setting of a temporal boundary. This viewpoint goes back to Mendon\c{c}a and co-workers in the early quantum literature on time refraction, where sudden temporal boundaries were shown to generate photon-pair creation through Bogoliubov mixing~\cite{mendoncca2000quantum,mendoncca2005time,mendoncca2011time}. Finite sequences of temporal slabs were then shown to realize temporal beam splitting and temporal interference~\cite{mendoncca2003temporal}. More recent quantum work has extended this picture to arbitrary temporal profiles, anisotropic temporal boundaries, temporal antireflection strategies, quantum state engineering, photon statistics at electromagnetic temporal interfaces, and, in specific time-reflection models, vacuum-entanglement effects~\cite{ganfornina2024quantum,vazquez2023shaping,stevens2025photon,liberal2023quantum,PhysRevResearch.7.013120_2025,svidzinsky2024time}. These time-reflection and time-refraction concepts have since reached beyond photonics, including recent demonstrations with ultracold atoms~\cite{dong2024quantum}.

\begin{figure*}[t]
    \centering
    \includegraphics[width=0.95\textwidth]{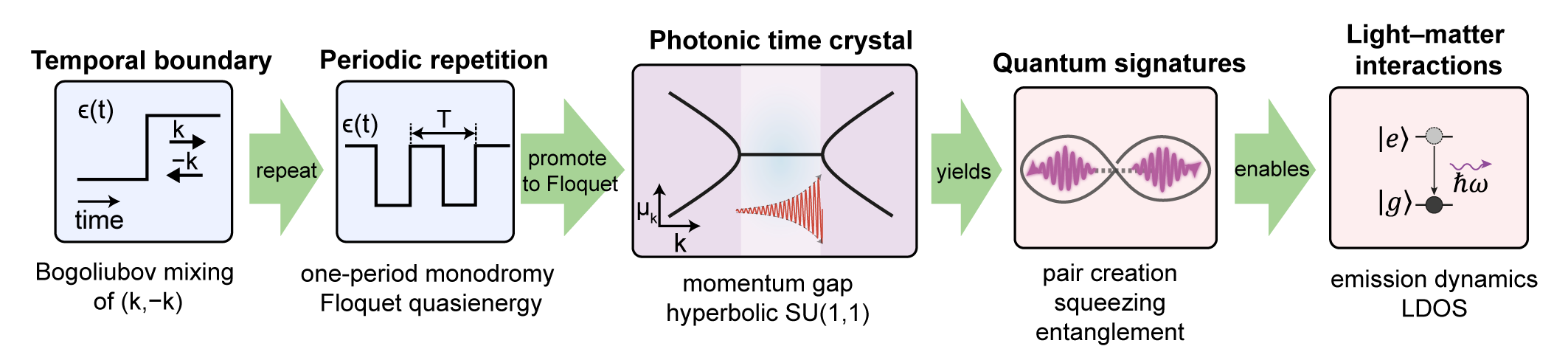}
    \caption{
Conceptual hierarchy of quantum PTCs. A single temporal boundary acts as an elementary Bogoliubov scatterer that mixes the \((k,-k)\) sector. Periodic repetition promotes this boundary-level process to a Floquet problem, giving rise to quasifrequency bands, momentum gaps, and hyperbolic \(SU(1,1)\) amplification in the momentum gap regime. {These structures underlie characteristic quantum signatures of idealized PTC models, including pair creation, squeezing, and nonclassical correlations, and motivate light--matter settings in which emission dynamics, LDOS signatures, and realistic platforms become directly relevant.}
}
    \label{fig:intro}
\end{figure*}

By contrast, systematic quantum treatments formulated specifically for the electromagnetic field in PTCs, in their band and momentum-gap regimes rather than in the broader DCE and nonstationary-media literature, remain comparatively recent. Important QED-oriented precursors had already highlighted momentum-gap-enhanced emission, lasing-like behavior, and free-electron radiation, including photon-pair creation from vacuum fluctuations in the gap regime~\cite{lyubarov2022amplified, dikopoltsev2022light}. A complementary line of work addressed vacuum amplification and photon-pair creation in PTCs built from periodically repeated abrupt temporal slabs, yielding an analytic theory that ties together transfer-matrix scattering, Bogoliubov mode mixing, two-mode squeezing, and DCE-like pair generation~\cite{sustaeta2025quantum}. Another quantum-electrodynamical treatment argued that, in an idealized Floquet-photonic synthetic-space formulation, the band/momentum-gap structure of a PTC can be read as a localization--delocalization transition: the momentum-gap regime corresponds to the delocalized side and the band regime to the localized side, a viewpoint that yields unconventional atom--photon dynamics~\cite{bae2025quantum}. Two adjacent directions sharpen the layer of this hierarchy: continuous-variable cluster-state generation has been proposed in PTCs~\cite{lyubarov2025continuous}, and a recent quantum-geometric formulation casts PTC mode evolution on the \(SU(1,1)\) coherent-state manifold~\cite{eswaran2025aspectsquantumgeometryphotonic}. Recent work further proposes, within a Floquet-BdG framework, that the Petermann factor of an effective PTC dynamical matrix can be tied to Bogoliubov mixing and to the scale of quantum squeezing and noise~\cite{kim2026petermann}.

These developments motivate the present review, which is deliberately focused rather than a comprehensive survey of all time-varying photonics. Its organizing thread is the quantum structure of PTCs and its link to light--matter interaction observables, running from temporal boundaries to Floquet media and, ultimately, to emitter- and LDOS-related settings. In Sec.~II, we begin with temporal boundaries as the elementary setting for Bogoliubov mode mixing and photon-pair creation in homogeneous \((k,-k)\) sectors, and then contrast this bulk picture with multimode mixing in spatially inhomogeneous cavities. In Sec.~III, we show how temporal periodicity promotes this temporal-boundary picture to a momentum-resolved Floquet theory of PTCs, with distinct band and momentum-gap regimes, a fixed-Nambu-basis formulation, and characteristic quantum observables. In Sec.~IV, we compare quantum PTCs with related frameworks for vacuum amplification, including the DCE, parametric amplification, \(SU(1,1)\) dynamics, and moving-grating or synthetic-motion viewpoints in spacetime-modulated QED. In Sec.~V, we turn to light--matter settings, reviewing spontaneous-emission decay and excitation, synthetic-space atom--photon dynamics, LDOS-based observables and the caveats in their interpretation, and broader experimental, material, and dispersive extensions. Figure~\ref{fig:intro} summarizes the core homogeneous hierarchy underlying the review.

Throughout the homogeneous bulk parts of Secs.~II and~III, \(\mathbf{k}\) denotes the wavevector and \(k\equiv |\mathbf{k}|\) its magnitude. Modes are labeled by \((\mathbf{k},\sigma)\), with \(\sigma\) denoting polarization; for the isotropic settings considered here, a real linear-polarization basis may be chosen, and the polarization index is suppressed whenever it plays no explicit role. Under these assumptions, the quadratic dynamics closes independently within each counter-propagating \((\mathbf{k},-\mathbf{k})\) pair. When the propagation direction is fixed or otherwise understood, we therefore suppress the vector notation and use the scalar label \(k\) as a shorthand for the corresponding \((\mathbf{k},-\mathbf{k})\) sector. In Sec.~\ref{sec:2-b}, by contrast, the spatially inhomogeneous cavity problem is naturally described by a discrete mode index \(n\), since the wavevector is no longer a good quantum number. 

\section{Temporal boundary}
\label{Sec2}

The PTCs considered in this focused review combine two distinct ingredients: temporal modulation and temporal periodicity. To separate their physical roles, we begin with the simplest nonperiodic setting, a single abrupt temporal boundary. In the idealized homogeneous, nondispersive setting used below, this elementary problem isolates positive--negative-frequency mixing, which becomes a Bogoliubov transformation and photon-pair creation once in/out or reference mode bases are chosen. The discussion is partly pedagogical, but it is useful: it exposes, in the simplest possible setting, the two-mode Bogoliubov structure that reappears later in homogeneous bulk PTC models.

We examine temporal mode mixing in two representative settings. Rather than starting from a fully general canonical quantization of arbitrary time-dependent media, we organize the discussion around Maxwell temporal boundary conditions. In a spatially homogeneous, nondispersive medium, this makes transparent why a temporal discontinuity closes the dynamics within each \((k,-k)\) sector and induces a two-mode Bogoliubov transformation. We then contrast this bulk picture with a spatially inhomogeneous cavity, where the natural basis is no longer a momentum basis and a temporal jump generally induces multimode mixing. The closed two-mode \(SU(1,1)\) structure of the homogeneous problem then serves throughout this section as a benchmark against which more general temporal quenches can be compared~\cite{mendoncca2000quantum,PhysRevResearch.7.013120_2025,PhysRevA.55.62}.

\subsection{Temporal boundaries and mode mixing}

We begin with an abrupt temporal-boundary setting, close to the early quantum treatment of time refraction by Mendon\c{c}a \textit{et al.}~\cite{mendoncca2000quantum}. The medium is taken to be homogeneous, nondispersive, nondissipative, and nonmagnetic, with a permittivity that changes suddenly at \(t=0\). Following the notation used throughout this review, we reserve \(\epsilon_0\) for the vacuum permittivity and write the pre- and post-jump absolute permittivities as \(\epsilon^{(0)}\) and \(\epsilon^{(1)}\), where the superscripts label stationary media:
\begin{equation}
\epsilon(t)=\epsilon^{(0)}\Theta(-t)+\epsilon^{(1)}\Theta(t),
\qquad
\epsilon^{(0)}\neq\epsilon^{(1)}.
\label{eq:single_jump_eps}
\end{equation}
To expose the operator-level mixing, we combine the standard temporal boundary conditions with a quantized plane-wave expansion. This is not meant as a fully general field-theoretic quantization of arbitrary time-dependent media; rather, within the present ideal homogeneous, nondispersive model it reproduces the same \((\mathbf{k},-\mathbf{k})\)-sector Bogoliubov structure found in more formal approaches. In a transverse plane-wave basis consistent with the Coulomb gauge, \(\nabla\cdot\mathbf A=0\), the electric field may be written as
\begin{equation}
\begin{aligned}
\mathbf E^{(j)}(\mathbf r,t)
&=
i\sum_{\mathbf{k},\sigma}
\sqrt{\frac{\hbar\omega_{k}^{(j)}}{2\epsilon^{(j)}}}
\Bigl[
a_{\mathbf{k}\sigma}^{(j)}
\mathbf u_{\mathbf{k}\sigma}(\mathbf r)e^{-i\omega_{k}^{(j)}t}
\\
&\qquad\qquad
-
{a_{\mathbf{k}\sigma}^{(j)}}^\dagger
\mathbf u_{\mathbf{k}\sigma}^*(\mathbf r)e^{i\omega_{k}^{(j)}t}
\Bigr],
\end{aligned}
\label{Efield}
\end{equation}
where \(j\in\{0,1\}\) labels the stationary media before and after the temporal jump, \(\epsilon^{(j)}\) is the corresponding absolute permittivity, and
\[
\mathbf u_{\mathbf{k}\sigma}(\mathbf r)
=
(2\pi)^{-3/2}e^{i\mathbf{k}\cdot\mathbf r}\hat{\mathbf e}_{\mathbf{k}\sigma}.
\]
Here and below, the summation notation is a compact shorthand for the mode expansion. Strictly, one should use either a finite-volume box normalization, \(\mathbf u_{\mathbf{k}\sigma}=V^{-1/2}e^{i\mathbf{k}\cdot\mathbf r}\hat{\mathbf e}_{\mathbf{k}\sigma}\), with discrete sums, or a continuum normalization with \(\sum_{\mathbf k}\) replaced by \(\int d^3 k\). The temporal matching and the resulting Bogoliubov coefficients do not depend on this bookkeeping convention.
The annihilation and creation operators satisfy the usual bosonic commutation relations. In Eq.~\eqref{Efield}, the field for a given \(\mathbf{k}\) is written as the sum of its positive- and negative-frequency parts. We adopt a real linear-polarization basis, for which \(\mathbf u_{\mathbf k\sigma}^{*}=\mathbf u_{-\mathbf k\sigma}\), up to the corresponding polarization-label convention.

The dispersion relation and the operator matching across the temporal boundary follow from the temporal boundary conditions at \(t=0\). In the absence of free charges, free currents, or singular source terms localized at the boundary, the standard ideal abrupt-switch model gives
\begin{equation}
\begin{aligned}
\mathbf{D}(\mathbf r,0^-)&=\mathbf{D}(\mathbf r,0^+),\\
\mathbf{B}(\mathbf r,0^-)&=\mathbf{B}(\mathbf r,0^+),
\end{aligned}
\label{Boundary_maxwell}
\end{equation}
with \(\mathbf D^{(j)}(\mathbf r,t)=\epsilon^{(j)}\mathbf E^{(j)}(\mathbf r,t)\). We stress that Eq.~\eqref{Boundary_maxwell} is used here only within this standard, externally imposed abrupt-switch model. More microscopic analyses of temporal boundaries have shown that the appropriate boundary conditions and conservation laws can depend on how the material parameters are actually varied in time~\cite{galiffi2025electrodynamics}. This qualification is worth keeping explicit: in moving- or time-boundary electrodynamics, boundary conditions acquire physical meaning only together with the material equations, and can depend on boundary-layer structure, material response times, and boundary velocity. The nondispersive jump model used here should therefore not be read as a universal microscopic boundary law~\cite{ostrovskii1975moving,galiffi2025electrodynamics}.

For a spatially uniform temporal jump, the boundary conditions must hold for all \(\mathbf r\), so the amplitudes multiplying \(e^{i\mathbf{k}\cdot\mathbf r}\) and \(e^{-i\mathbf{k}\cdot\mathbf r}\) are matched mode by mode. The wavevector \(\mathbf k\) is therefore conserved, while the frequency is set separately in each stationary region by the corresponding dispersion relation,
\[
\omega_k^{(j)}=\frac{c|\mathbf k|}{n^{(j)}} ,
\]
with \(n^{(j)}\) the refractive index of the \(j\)th medium. The temporal discontinuity then mixes positive- and negative-frequency components within each fixed \((\mathbf{k},-\mathbf{k})\) sector. Using \(\mathbf u_{\mathbf{k}\sigma}^{*}=\mathbf u_{-\mathbf{k}\sigma}\), the matching conditions yield
\begin{equation}
\begin{aligned}
\sqrt{\frac{\epsilon^{(0)}}{n^{(0)}}}
\left(a_{\mathbf{k}}^{(0)}-{a_{-\mathbf{k}}^{(0)}}^\dagger\right)
&=
\sqrt{\frac{\epsilon^{(1)}}{n^{(1)}}}
\left(a_{\mathbf{k}}^{(1)}-{a_{-\mathbf{k}}^{(1)}}^\dagger\right),
\\
\sqrt{\frac{n^{(0)}}{\epsilon^{(0)}}}
\left(a_{\mathbf{k}}^{(0)}+{a_{-\mathbf{k}}^{(0)}}^\dagger\right)
&=
\sqrt{\frac{n^{(1)}}{\epsilon^{(1)}}}
\left(a_{\mathbf{k}}^{(1)}+{a_{-\mathbf{k}}^{(1)}}^\dagger\right),
\end{aligned}
\label{eq:temporal_matching_operators}
\end{equation}
where the two relations follow from the continuity of \(\mathbf D\) and \(\mathbf B\), respectively. Rearranging gives the Bogoliubov relation
\begin{equation}
\begin{pmatrix}
a_{\mathbf{k}}^{(1)}(0^+)\\
{a_{-\mathbf{k}}^{(1)}}^\dagger(0^+)
\end{pmatrix}
=
\begin{pmatrix}
A & B\\
B & A
\end{pmatrix}
\begin{pmatrix}
a_{\mathbf{k}}^{(0)}(0^-)\\
{a_{-\mathbf{k}}^{(0)}}^\dagger(0^-)
\end{pmatrix},
\label{Bogoliubov}
\end{equation}
where
\[
A=\frac{1+\alpha^2}{2\alpha},
\qquad
B=\frac{1-\alpha^2}{2\alpha},
\qquad
\alpha=\sqrt{\frac{n^{(0)}}{n^{(1)}}}.
\]
These coefficients satisfy \(A^2-B^2=1\), so the transformation preserves the bosonic commutation relations.

For the fixed homogeneous \((\mathbf{k},-\mathbf{k})\) sector under consideration, we now suppress the vector notation and denote the sector simply by \(k\). Because the sudden jump changes the stationary normal-mode basis, it is natural to introduce distinct annihilation and creation operators before and after the boundary~\cite{mendoncca2000quantum,law1994effective}. Photon number is therefore basis-dependent in this time-dependent problem: the temporal jump relates two stationary Fock bases, so photon occupation must be defined relative to a specified reference mode basis. We label the corresponding Fock states by the temporal-region index \(j=0,1\), writing \(\ket{n_k,n_{-k}}_j\). Using Eq.~\eqref{Bogoliubov}, a pre-jump Fock state can be expanded in the post-jump basis as
\begin{equation}
\ket{n,0}_{0}
=
\sum_{s=0}^{\infty}
b_s(n)\ket{n+s,s}_{1},
\label{state}
\end{equation}
where
\[
b_s(n)=
\sqrt{1-(B/A)^2}
\sqrt{\frac{(s+n)!}{s!\,n!}}\,
B^s A^{-(n+s)}.
\]
Equation~\eqref{state} is a basis-transformation identity between the Fock bases of the two stationary media, not a statement about dynamical evolution. For an initial vacuum \(\ket{0,0}_0\), the same state expressed immediately after the jump in the post-jump basis takes the form
\[
\ket{\psi(0^+)}=\sum_s b_s(0)\ket{s,s}_{1},
\]
that is, a two-mode squeezed superposition in the \(1\)-basis. Since the medium is again stationary for \(t>0\), the subsequent evolution is governed by the post-jump Hamiltonian, equivalently by the post-jump mode operators \(a_{\pm k}^{(1)}\) oscillating at frequency \(\omega_k^{(1)}\).

It is often useful to re-express this post-jump evolution in the fixed pre-jump basis. In this fixed-basis description, pair creation and the corresponding mode occupations can be tracked directly relative to the initial vacuum, while the underlying \(SU(1,1)\) structure of the dynamics becomes explicit. Using \(A^2-B^2=1\), the inverse of Eq.~\eqref{Bogoliubov} reads
\begin{equation}
a_{k}^{(0)}
=
A\,a_{k}^{(1)}
-
B\,{a_{-k}^{(1)}}^\dagger .
\label{inverse_Bogoliubov}
\end{equation}
For \(t>0\),
\begin{equation}
\begin{aligned}
a_{k}^{(1)}(t)&=a_{k}^{(1)}e^{-i\omega_{k}^{(1)}t},\\
{a_{-k}^{(1)}}^\dagger(t)&={a_{-k}^{(1)}}^\dagger e^{+i\omega_{k}^{(1)}t}.
\end{aligned}
\label{post_jump_evolution}
\end{equation}
The \(0\)-basis operator evolved under the post-jump Hamiltonian is therefore
\begin{equation}
a_{k}^{(0)}(t)
=
A\,a_{k}^{(1)} e^{-i\omega_{k}^{(1)}t}
-
B\,{a_{-k}^{(1)}}^\dagger e^{i\omega_{k}^{(1)}t}.
\end{equation}
Substituting the forward relation, Eq.~\eqref{Bogoliubov}, gives
\begin{equation}
a_{k}^{(0)}(t)
=
u_k(t)a_{k}^{(0)}
+
v_k(t){a_{-k}^{(0)}}^\dagger,
\end{equation}
with
\[
\begin{aligned}
u_k(t)&=A^2 e^{-i\omega_{k}^{(1)}t}-B^2 e^{i\omega_{k}^{(1)}t},\\
v_k(t)&=-2iAB\sin\!\bigl(\omega_{k}^{(1)}t\bigr).
\end{aligned}
\]

In terms of the Heisenberg-evolved Nambu spinor in the fixed \(0\)-basis, the post-jump dynamics takes the form
\begin{equation}
\begin{aligned}
\begin{pmatrix}
a_{k}^{(0)}(t)\\
{a_{-k}^{(0)}}^\dagger(t)
\end{pmatrix}
&=
\begin{pmatrix}
u_k(t) & v_k(t)\\
v_k^*(t) & u_k^*(t)
\end{pmatrix}
\begin{pmatrix}
a_{k}^{(0)}\\
{a_{-k}^{(0)}}^\dagger
\end{pmatrix}
\\
&\equiv
\mathcal U_k(t,0)
\begin{pmatrix}
a_{k}^{(0)}\\
{a_{-k}^{(0)}}^\dagger
\end{pmatrix}.
\end{aligned}
\label{eq:fixedbasis}
\end{equation}
The Nambu-space propagator \(\mathcal U_k(t,0)\) obeys the standard two-mode Bogoliubov constraint
\[
|u_k(t)|^2-|v_k(t)|^2=1,
\]
consistent with an \(SU(1,1)\) Bogoliubov evolution in the two-mode Nambu space. For the state evolved from an initial vacuum in the \(0\)-basis, the probability of finding \(n\) photons in each of the two modes is
\begin{equation}
P^{(0)}_{n,n}(t)
=
\frac{|v_k(t)|^{2n}}{|u_k(t)|^{2n+2}}.
\end{equation}
Thus, at a homogeneous temporal boundary, spatial translational symmetry closes the dynamics within each \((k,-k)\) block and produces the \(SU(1,1)\) structure underlying squeezing and photon-pair creation. In this fixed-basis convention, the Bogoliubov coefficients fix the photon-pair probabilities relative to the chosen initial vacuum. The same logic returns in Sec.~\ref{sec:3-2}, where periodically driven PTCs are quantized in a fixed reference basis so that occupations, pair correlations, and squeezing measures can all be expressed relative to a specified set of reference modes.

\subsection{Multimode Bogoliubov mixing beyond homogeneous media}
\label{sec:2-b}

\begin{figure}[t]
    \centering
    \includegraphics[width=0.65\columnwidth]{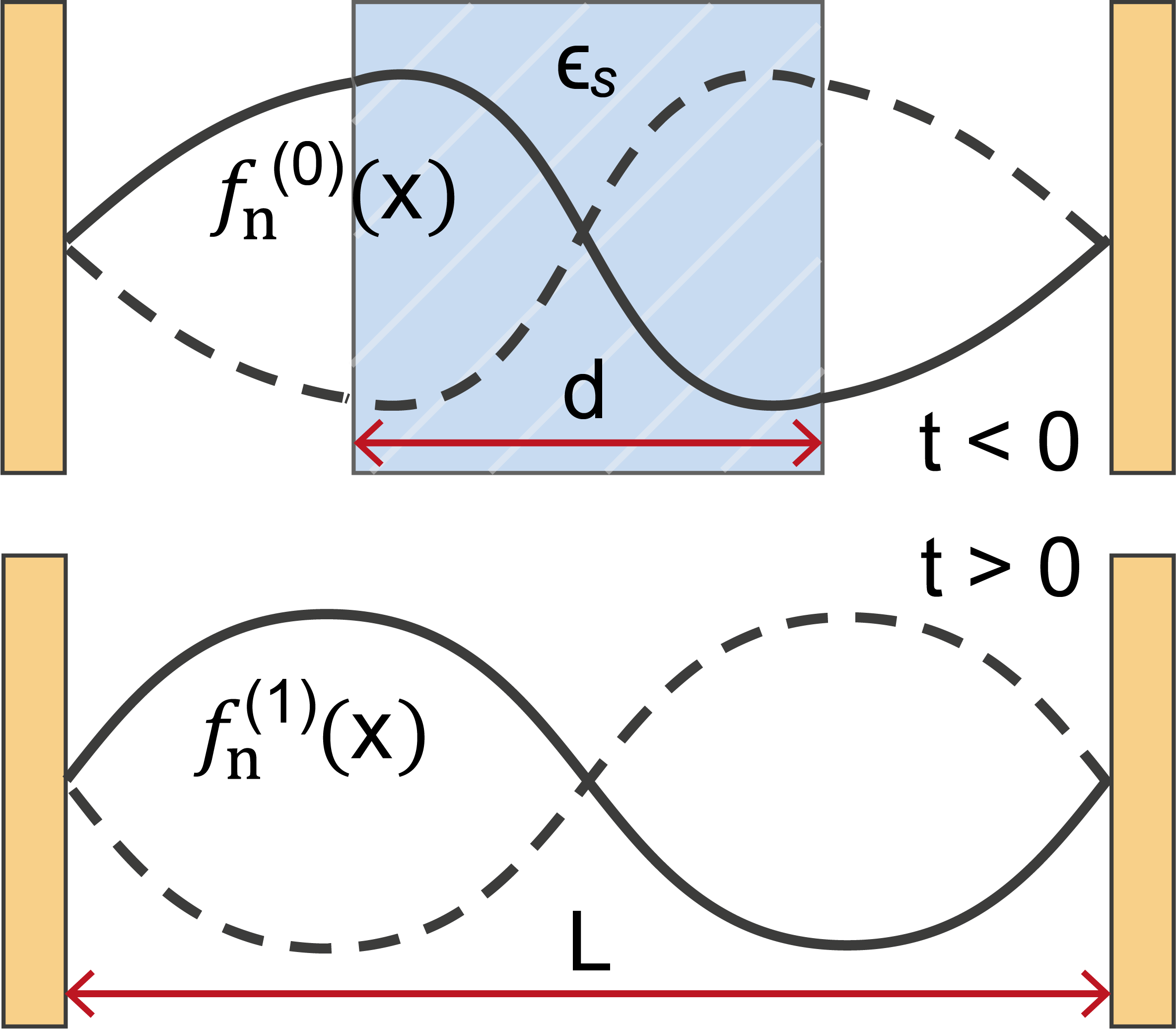}
    \caption{
Schematic illustration of multimode mixing induced by sudden dielectric removal from a cavity.
Top: for \(t<0\), a dielectric slab of width \(d\) and permittivity \(\epsilon_s\) sits inside a cavity of length \(L\), and the field is expanded in the corresponding inhomogeneous cavity modes \(\{f_n^{(0)}(x)\}\).
Bottom: for \(t>0\), the slab is removed and the cavity becomes empty, so the field must be re-expanded in the empty-cavity modes \(\{f_n^{(1)}(x)\}\).
Because the two configurations are governed by different boundary-value problems, a single initial mode generally has nonzero projections onto several final modes, producing multimode Bogoliubov mixing.}
\label{fig:cavity}
\end{figure}

A useful contrast with the homogeneous temporal-boundary problem comes from closed, lossless, spatially inhomogeneous cavities. We use this setting not as a full cavity-QED derivation, but as a focused example of how the natural mode basis changes when translational symmetry is lost. The clean two-mode \((k,-k)\) closure of the homogeneous bulk problem is special: once translational symmetry is broken, a temporal quench generally couples distinct cavity eigenmodes rather than preserving an isolated momentum sector. Early cavity-based studies of time-dependent boundaries and nonstationary dielectric media already pointed in this direction~\cite{law1994effective,PhysRevA.53.1031}, and a particularly explicit soluble example was given by Cirone \textit{et al.}~\cite{PhysRevA.55.62}. There, an effectively one-dimensional cavity containing a dielectric slab undergoes a sudden temporal change at \(t=0\), after which the field evolves as that of an empty cavity, as sketched in Fig.~\ref{fig:cavity}.

The difference is not merely notational because the natural basis differs qualitatively from that of the homogeneous bulk problem. In a homogeneous medium, translational symmetry preserves the wavevector, so the temporal matching closes within each \((k,-k)\) sector. In a cavity or other spatially inhomogeneous setting, the appropriate description is instead in terms of discrete normal modes set by the corresponding boundary-value problem and its weighted inner product~\cite{knight_PhysRevA.54.2292}. Temporal matching then generally induces intermode Bogoliubov mixing rather than the isolated \(SU(1,1)\) block of the homogeneous case. Exceptions arise when the spatial eigenbasis stays fixed in time; in factorized media where the spatial and temporal dependences separate, for instance, the dynamics can reduce to an effectively mode-diagonal form~\cite{dodonov_PhysRevA.47.4422}.

For the field description, one may again work in the Coulomb gauge and, in the present one-dimensional single-polarization reduction, expand the relevant component of the vector potential as
\begin{equation}
A^{(j)}(x,t)
=
\sum_n
\left[
\alpha_n^{(j)} f_n^{(j)}(x)e^{-i\omega_n^{(j)}t}
+\text{c.c.}
\right],
\end{equation}
where \(j=0,1\) labels the stationary cavity configurations before and after the temporal boundary, \(f_n^{(j)}(x)\) is the spatial mode profile, and \(\alpha_n^{(j)}\) the corresponding complex mode amplitude. Because the dielectric profile is spatially inhomogeneous, translational symmetry is broken and the wavevector is no longer a good quantum number; the field is instead decomposed into the normal modes of the corresponding boundary-value problem. In the present nonmagnetic reduction, these modes satisfy
\[
\left[
\partial_x^2+
\left(\frac{\omega_n^{(j)}}{c}\right)^2
\frac{\epsilon^{(j)}(x)}{\epsilon_0}
\right]
f_n^{(j)}(x)=0,
\]
together with the appropriate cavity boundary conditions and slab-interface matching. The expansion is written here at the level of classical mode amplitudes for clarity; upon quantization, the properly normalized modal amplitudes are promoted to ladder operators associated with the same normal modes. Since the boundary-value problem is real and lossless, the mode functions may be chosen real up to an overall phase.

The cavity walls are assumed to be perfect electric conductors. Since the tangential electric field vanishes at the walls and \(E=-\partial_t A\), the mode functions obey \(f_n^{(j)}(x)=0\) there. At the slab--vacuum interfaces, continuity of the tangential electric and magnetic fields implies, in this one-dimensional nonmagnetic reduction, continuity of \(f_n^{(j)}(x)\) and of its first spatial derivative. Within each stationary cavity configuration, the modes may be normalized with the dielectric-weighted inner product
\begin{equation}
\langle f_m^{(j)}|f_n^{(j)}\rangle_{\epsilon^{(j)}}
\equiv
\int dx\,\epsilon^{(j)}(x)\,
f_m^{(j)}(x)^*
f_n^{(j)}(x),
\end{equation}
so that \(\langle f_m^{(j)}|f_n^{(j)}\rangle_{\epsilon^{(j)}}=\delta_{mn}\) under the chosen normalization convention~\cite{knight_PhysRevA.54.2292}.

Having determined \(f_n^{(j)}(x)\) and \(\omega_n^{(j)}\) in the two stationary cavity configurations, one imposes the temporal matching conditions at \(t=0\) to relate the corresponding mode amplitudes and, after quantization, the associated operators. When the slab is suddenly removed, the field is re-expanded in the empty-cavity modes, a family generally distinct from that of the dielectric-loaded cavity. A single initial mode then has nonzero projections onto several final modes, and the temporal matching cannot be reduced to a single-mode correspondence. The resulting operator transformation has the multimode Bogoliubov form
\begin{equation}
\begin{aligned}
a_n^{(1)}
&=
\sum_m
\left(
h_{n,m}a_m^{(0)}
+
g_{n,m}{a_m^{(0)}}^\dagger
\right),
\\
{a_n^{(1)}}^\dagger
&=
\sum_m
\left(
h_{n,m}^*{a_m^{(0)}}^\dagger
+
g_{n,m}^*a_m^{(0)}
\right),
\end{aligned}
\end{equation}
where \(a_n^{(0)}\) and \(a_n^{(1)}\) are the annihilation operators of the cavity-mode families before and after the temporal boundary, respectively, satisfying
\[
[a_n^{(0)},{a_{n'}^{(0)}}^\dagger]=\delta_{nn'},
\qquad
[a_n^{(1)},{a_{n'}^{(1)}}^\dagger]=\delta_{nn'}.
\]
Preservation of these commutation relations imposes the usual multimode symplectic constraints on the Bogoliubov coefficients: the matrices \(h\) and \(g\) must satisfy relations of the form \(hh^\dagger-gg^\dagger=I\) and \(hg^{\mathsf T}-gh^{\mathsf T}=0\), with indices arranged according to the convention used in the transformation above.

The homogeneous and inhomogeneous cases therefore differ not in whether Bogoliubov mixing occurs, but in how it is organized. In the homogeneous bulk problem, spatial translational symmetry enforces a momentum-sector decomposition and reduces the temporal boundary to independent two-mode \((k,-k)\) blocks. In the cavity problem, such a decomposition is no longer natural; the natural basis is instead a set of discrete normal modes, and the same temporal matching typically produces a multimode Bogoliubov transformation. This is the main lesson of the section: the two-mode \(SU(1,1)\) block is the elementary, highly transparent homogeneous limit, whereas spatial inhomogeneity generally promotes temporal-boundary mixing to a multimode problem.

\section{Quantum theory of photonic time crystals}\label{sec:3}

PTCs may be viewed as periodic continuations of the temporal-boundary physics discussed above, but their distinguishing feature is that temporal periodicity promotes each conserved-momentum sector to a Floquet problem over one modulation period. The resulting Floquet modes and quasifrequencies define the band structure of the PTC, including its band and momentum-gap regimes. Classically, this structure is usually obtained from transfer-matrix or Maxwell--Floquet formulations, whose one-period maps or dynamical matrices can be cast as effective non-Hermitian descriptions of the quasifrequency spectrum. In this language, mode nonorthogonality and amplification in the momentum gap already appear within classical wave theory, while related LDOS peaks and emitter-response features can often be captured at a classical or semiclassical level. Exceptional points can also arise at gap edges or in suitable parameter regimes within these effective Floquet descriptions~\cite{lustig2018topological,PhysRevB.98.085142,gaxiola2021temporal,park2022revealing,asgari2024theory}.

The quantum theory builds on this classical backbone but targets a distinct set of observables. Quantum signatures such as photon-pair creation, two-mode squeezing, and anomalous or nonclassical correlations between counter-propagating modes go beyond classical field amplification: they are operator-level consequences of positive--negative-frequency mixing in a time-dependent medium.
Although the classical electrodynamics of PTCs is by now well developed, quantum treatments aimed specifically at PTC band and momentum-gap physics have only recently begun to be organized systematically. Historically, important precursors already existed in the broader literature on quantum electrodynamics in time-dependent media. Dodonov \textit{et al.}, in particular, developed an early quantization framework for nonstationary linear media and obtained explicit results for factorized time-dependent settings, though not yet in the modern Floquet-band language~\cite{dodonov_PhysRevA.47.4422}. A distinct but complementary line is the temporal-slab literature, where finite sequences of temporal jumps, temporal interference, and Bogoliubov mixing were developed in forms tied more directly to the slab-based quantum route later used for PTCs~\cite{mendoncca2003temporal}.

Recent quantum work on PTCs splits naturally into two closely related threads. One line treats the time-modulated permittivity as a prescribed periodic background and quantizes the electromagnetic field, focusing on photon-pair creation, squeezing, and vacuum amplification in idealized homogeneous PTC models~\cite{sustaeta2025quantum}. This approach makes explicit the link between transfer-matrix scattering, Bogoliubov mode mixing, two-mode squeezing, and DCE-like pair generation. A complementary quantum-electrodynamical line couples the PTC field to an explicit atomic degree of freedom and argues, within a Floquet-photonic synthetic-space formulation, that the band/momentum-gap structure of the bulk homogeneous PTC can be read in terms of localization--delocalization behavior and can give rise to unconventional atom--photon dynamics~\cite{bae2025quantum}. Together, these frameworks clarify how quantum observables intrinsic to idealized PTC models can be defined, especially in and near the momentum-gap regime, and how effective non-Hermitian classical descriptions relate to an underlying Hermitian quadratic quantum dynamics~\cite{nonH2019PhysRevA.99.063834,bae2025quantum}. Throughout this review, ``non-Hermitian'' should accordingly be read at the level of the relevant Maxwell--Floquet, transfer-matrix, or bosonic BdG dynamical representation, not automatically as an open-system dissipative Hamiltonian. In ideal lossless PTC models, complex Floquet exponents describe parametric amplification driven by external temporal modulation; genuine absorption, material gain, or reservoir-induced quantum noise requires extra bath degrees of freedom and a corresponding open-system description~\cite{ashida2020nonhermitian}.

In this section, we first review the classical Floquet theory generated by temporal periodicity and the resulting band and momentum-gap structure. We then turn to a quantum description of PTCs based on a canonical formulation for prescribed temporal profiles; imposing periodicity leads to the Floquet problem relevant for ideal PTC models. We show how this formulation can be recast as Heisenberg dynamics in a fixed Nambu basis, which makes explicit its continuity with the temporal-boundary physics discussed above. Finally, within this quantum framework, we discuss how the effective non-Hermitian features of the classical Floquet representation are reflected in closed quantum BdG dynamics, and we examine representative observables such as photon number, squeezing, and pair correlations defined relative to specified reference or output modes.

\subsection{Temporal periodicity, monodromy, and Floquet band structure}\label{3-1}

To connect the temporal-boundary discussion of Sec.~\ref{Sec2} to the Floquet band structure of PTCs, we first consider the classical problem in a homogeneous, nondispersive, nondissipative medium. In this ideal lossless bulk model, temporal periodicity promotes each conserved-momentum sector to a one-period monodromy problem. A nonzero imaginary part of the Floquet quasifrequency then marks the parametrically amplified or attenuated regime, which we identify with the PTC momentum gap. In lossy, dispersive, finite-size, or reservoir-coupled settings, however, complex quasifrequencies can also encode material damping, radiative leakage, or reservoir-induced gain and loss, so we tie their interpretation to the specific dynamical model at hand.
The same quasifrequency spectrum can also be obtained by recasting Maxwell's equations as a first-order time-evolution problem or by using plane-wave Floquet formulations. We adopt the transfer-matrix approach because it stays most directly continuous with the temporal-boundary matching developed in Sec.~\ref{Sec2}.

For the stepwise construction below, we use a temporal boundary-matching formulation, closely related to the Kr\"onig--Penney treatment of square-profile temporal photonic crystals~\cite{gaxiola2021temporal}, which we cast as a one-period transfer, or monodromy, matrix. This is the same transfer-matrix viewpoint adopted in more general treatments of temporally stratified media~\cite{asgari2024theory,xu2022generalized}, while the broader PTC band picture traces back to the dynamic-slab Bloch--Floquet formulation of Ref.~\cite{zurita2009reflection}. In direct analogy with the single-boundary problem of Sec.~\ref{Sec2}, the field in each temporal segment is described by forward- and backward-propagating components, and we impose the temporal boundary conditions at each switching time. Because the medium remains spatially homogeneous, the matching closes independently within each conserved-momentum sector, acting on the coefficients of the spatial phase factors \(e^{\pm i\mathbf{k}\cdot\mathbf r}\). For concreteness, we take the propagation direction along the \(x\) axis, so that the wavevector label reduces to \(k\).
\begin{figure}[t]
    \centering
    \hspace*{-1.0cm}
    \includegraphics[width=0.8\columnwidth]{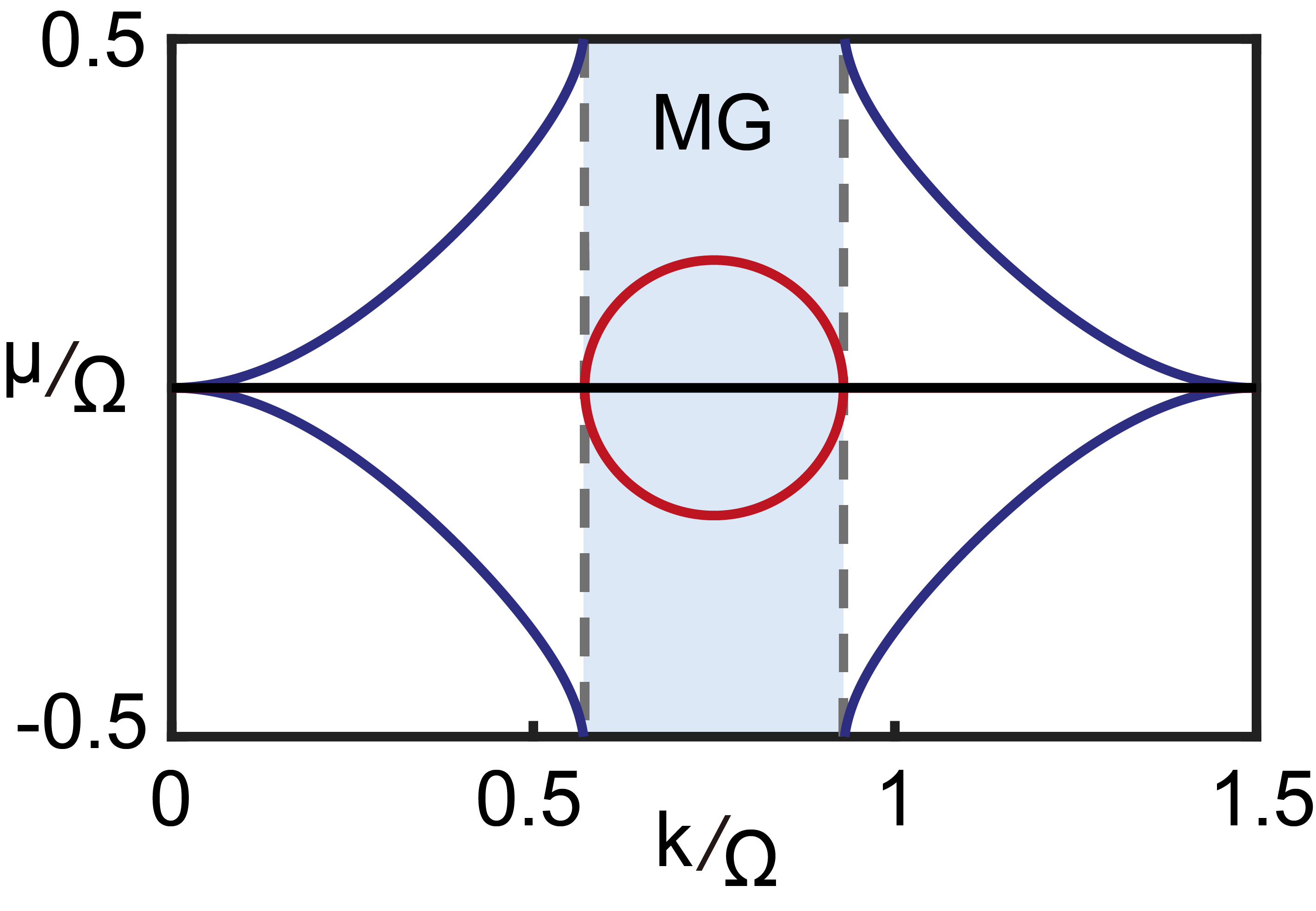}
    \caption{
Floquet quasifrequency spectrum of the stepwise PTC in Eq.~\eqref{slab}. Real (blue) and imaginary (red) parts of the Floquet quasifrequency \(\mu_k\) are plotted versus \(k/\Omega\), where \(\Omega=2\pi/T\). {In the lossless monodromy model shown here, the shaded momentum-gap regime, \(\mathrm{Im}(\mu_k)\neq 0\), marks the exponentially amplified and attenuated Floquet modes.} Here \(n^{(0)}=1.5\), \(n^{(1)}=0.5\), \(t_1=1\), \(T=2\), and \(c=1\).
}
    \label{fig:band}
\end{figure}

As a representative example, consider a stepwise PTC built by periodically repeating a two-step temporal slab of period \(T\),
\begin{equation}
\epsilon(t)=
\begin{cases}
\epsilon^{(1)}, & (\ell-1)T<t<\ell T-t_1,\\[2mm]
\epsilon^{(0)}, & \ell T-t_1<t<\ell T,
\end{cases}
\qquad \ell\in\mathbb{Z}.
\label{slab}
\end{equation}
Here the \(\ell\)th temporal unit cell is defined by \((\ell-1)T<t<\ell T\), and \(t_1\) is the duration of the \(\epsilon^{(0)}\) segment. {In this transfer-matrix subsection, \(\epsilon^{(j)}\) denotes the segment permittivity entering the classical temporal-boundary matching; in the fixed-basis canonical formulation below, we specify the relative permittivity explicitly.} For a fixed wavevector \(k\), we write the electric field in each temporal segment as
\begin{equation}
E_k^{(j)}(x,t)
=
\alpha_{\ell,k}^{(j)}e^{i(kx-\omega_k^{(j)}t)}
+\text{c.c.},
\end{equation}
where \(j=0,1\) refers to the segments with permittivities \(\epsilon^{(0)}\) and \(\epsilon^{(1)}\), respectively, and \(\ell\) labels the temporal unit cell. Spatial homogeneity preserves the wavevector, and each segment obeys
\[
\omega_k^{(j)}=\frac{c|k|}{n^{(j)}}.
\]
Equivalently, we encode the real-field expansion in the two-component amplitude vector
\[
\bigl(\alpha_{\ell,k}^{(j)},(\alpha_{\ell,-k}^{(j)})^*\bigr)^\mathsf{T},
\]
which is the classical analogue of the Nambu spinor used below.

Applying the temporal boundary conditions of Eq.~\eqref{Boundary_maxwell} at the two switching times, together with the phase accumulation in each temporal segment, yields the one-period relation
\begin{equation}
\begin{aligned}
\begin{pmatrix}
\alpha_{\ell,k}^{(0)}\\
(\alpha_{\ell,-k}^{(0)})^*
\end{pmatrix}
&=
\begin{pmatrix}
\mathcal{M}_{11} & \mathcal{M}_{12}\\
\mathcal{M}_{12}^* & \mathcal{M}_{11}^*
\end{pmatrix}
\begin{pmatrix}
\alpha_{\ell-1,k}^{(0)}\\
(\alpha_{\ell-1,-k}^{(0)})^*
\end{pmatrix}
\\
&\equiv
\mathcal{M}_k
\begin{pmatrix}
\alpha_{\ell-1,k}^{(0)}\\
(\alpha_{\ell-1,-k}^{(0)})^*
\end{pmatrix}.
\end{aligned}
\label{eq:monodromy}
\end{equation}
Here \(\mathcal{M}_k\) is the monodromy matrix over one period \(T\). Like the single-boundary Bogoliubov matrix of Sec.~\ref{Sec2}, it has a pseudo-unitary two-component structure and satisfies
\[
|\mathcal{M}_{11}|^2-|\mathcal{M}_{12}|^2=1.
\]
Temporal periodicity therefore promotes the elementary boundary-induced mixing of Sec.~\ref{Sec2} to a momentum-resolved one-period Floquet map.
The Floquet band structure follows from the eigenvalue problem of \(\mathcal{M}_k\). Writing the stroboscopic evolution as
\[
\mathbf{\Psi}_{\ell,k}=\mathcal{M}_k\,\mathbf{\Psi}_{\ell-1,k},
\]
the Floquet modes are the eigenvectors of \(\mathcal{M}_k\),
\begin{equation}
\mathbf{\Psi}_{\ell,k}
=
\lambda_k\,\mathbf{\Psi}_{\ell-1,k},
\end{equation}
where \(\lambda_k\) is an eigenvalue of the monodromy matrix.

For a system subject to periodic driving, Floquet theory implies that the corresponding Floquet solutions may be written in the form~\cite{Floquet1,Floquet2}
\begin{equation}
\mathbf{\Psi}_k(t)
=
\mathbf{u}_k(t)e^{-i\mu_k t},
\qquad
\mathbf{u}_k(t+T)=\mathbf{u}_k(t),
\label{Floquetstate}
\end{equation}
where \(\mu_k\) is the Floquet quasifrequency. {As is standard in Floquet problems, \(\mu_k\) is defined only modulo \(\Omega=2\pi/T\): choosing a branch of the logarithm fixes a temporal Brillouin zone, while the growth or decay rate is set by the modulus of \(\lambda_k\).} Stroboscopic evaluation at \(t=(\ell-1)T\) and \(t=\ell T\) gives
\[
\mathbf{\Psi}_k(\ell T)
=
e^{-i\mu_k T}
\mathbf{\Psi}_k((\ell-1)T),
\]
so
\begin{equation}
\lambda_k=e^{-i\mu_k T}.
\end{equation}
When \(\mu_k\) is real, the corresponding Floquet mode remains oscillatory in time. When \(\mu_k=\nu_k+i\gamma_k\), the factor \(e^{\gamma_k t}\) produces exponential amplification or attenuation. {For the ideal nondissipative PTC considered here, a nonzero imaginary part identifies the parametric momentum-gap regime. A representative Floquet quasifrequency spectrum is shown in Fig.~\ref{fig:band}. The real part of \(\mu_k\) forms the quasifrequency bands, and the shaded momentum-gap region is characterized by \(\mathrm{Im}(\mu_k)\neq0\). This classical Floquet structure is the direct precursor of the quantum amplification, fixed-basis Nambu dynamics, and observables developed in the following subsections.

\subsection{Canonical quantum formulation in a fixed Nambu basis}\label{sec:3-2}

Within the temporal-slab literature, the quantum precursor closest in spirit to later slab-based quantum descriptions of PTCs is the work of Mendon\c{c}a \textit{et al.}~\cite{mendoncca2003temporal}. There, temporal periodicity is not yet the primary organizing principle; the emphasis is on the quantum properties of temporal interference and temporal beam splitting generated by a finite sequence of time slabs. Sustaeta-Osuna \textit{et al.} instead considered a PTC formed by the periodic repetition of abrupt temporal slabs and elevated the corresponding transfer-matrix description to the level of quantized operators~\cite{sustaeta2025quantum}, providing a stroboscopic quantum description of photon-pair creation and amplification in a PTC. By extracting the Floquet quasifrequency from the eigenvalues of the transfer matrix, they connected momentum gaps to exponential quantum amplification and related classical reflectivity to quantum pair generation through the squeezing parameter.

The slab-based formalism gives analytic control, including in the momentum-gap regime, and keeps the classical-to-quantum correspondence transparent. Since each constant-permittivity temporal slab has its own diagonal mode basis, the evolution is represented as phase accumulation within each slab followed by Bogoliubov mixing at the switching times, making the formalism especially natural for constructing the one-period transfer matrix. To track particle creation relative to a single initial vacuum throughout the evolution, the same dynamics can be re-expressed in a fixed initial basis, as in Eq.~\eqref{eq:fixedbasis}, through the Bogoliubov relation induced by the transfer matrix. This motivates a complementary formulation in which the problem is posed directly in a fixed reference basis through canonical quantization. The present subsection is the periodic analogue of the fixed-basis Bogoliubov description introduced for a single temporal boundary in Sec.~\ref{Sec2}.

A useful route toward such a fixed-basis formulation was used in one early quantum-oriented treatment of PTCs by Lyubarov \textit{et al.}~\cite{lyubarov2022amplified}. For a homogeneous medium with time-dependent relative permittivity \(\epsilon(t)\), {so that the absolute permittivity is \(\varepsilon(t)=\epsilon_0\epsilon(t)\),} and fixed permeability \(\mu=\mu_0\), one may start from the classical Lagrangian density in the Coulomb gauge,
\[
\mathcal{L}(\mathbf{r},t)
=
\frac{\epsilon_0\epsilon(t)}{2}
(\partial_t \mathbf A)^2
-
\frac{1}{2\mu_0}
(\nabla\times \mathbf A)^2 .
\]
{With the convention \(\mathbf E=-\partial_t\mathbf A\), the canonical conjugate momentum of the vector potential is \(\mathbf{\Pi}=\partial\mathcal L/\partial(\partial_t\mathbf A)=-\mathbf{D}\).} In terms of the canonical variables, the Hamiltonian is
\begin{equation}
H
=
\frac{\epsilon_0}{2}
\int d^3r
\left[
\frac{\mathbf \Pi^2}{\epsilon_0^2\epsilon(t)}
+
c^2(\nabla\times \mathbf A)^2
\right].
\end{equation}
We then promote the canonical fields to operators and expand \(\hat{\mathbf A}(\mathbf r,t)\) in a time-independent plane-wave basis,
\begin{equation}
\hat{\mathbf A}(\mathbf r,t)
=
\frac{1}{\sqrt{\epsilon_0 V}}
\sum_{\mathbf k,\sigma}
\hat{\mathbf e}_{\mathbf k\sigma}
e^{i\mathbf k\cdot\mathbf r}
\hat q_{\mathbf k\sigma}(t),
\end{equation}
with the reality condition imposed when the full Hermitian field is reconstructed. {Here \(V\) denotes a finite quantization volume; the continuum limit is obtained by the usual replacement of discrete sums by integrals with the corresponding normalization.} In this representation, the canonical variables \(\hat q_{\mathbf k\sigma}\) and \(\hat p_{\mathbf k\sigma}\) are attached to basis functions that do not themselves vary in time. Expanding the conjugate momentum analogously in terms of \(\hat p_{\mathbf{k}\sigma}\) gives
\begin{equation}
H
=
\frac{1}{2}
\sum_{\mathbf{k}\sigma}
\left[
\frac{1}{\epsilon(t)}
\hat{p}_{\mathbf{k}\sigma}
\hat{p}_{\mathbf{k}\sigma}^\dagger
+
c^2k^2
\hat{q}_{\mathbf{k}\sigma}
\hat{q}_{\mathbf{k}\sigma}^\dagger
\right].
\end{equation}
A detailed derivation is given in the Supplementary Material of Ref.~\cite{lyubarov2022amplified}. {A related quantum treatment for time-varying photonic media with arbitrary temporal profiles of the permittivity and permeability was later introduced by Ganfornina-Andrades \textit{et al.}~\cite{ganfornina2024quantum}.}

At this stage, we introduce ladder operators for the resulting time-dependent harmonic-oscillator problem. Rather than a time-dependent operator definition adapted to an instantaneous mode basis, as in the approach of Law~\cite{law1994effective}, the fixed-basis formulation defines the operators with respect to a constant reference index \(n_r\), for example a temporal average of \(n(t)\):
\begin{equation}
\begin{aligned}
\hat a_{\mathbf k\sigma}
&=
\sqrt{\frac{ck n_r}{2\hbar}}\,
\hat q_{\mathbf k\sigma}
+
i\sqrt{\frac{1}{2\hbar ck n_r}}\,
\hat p^\dagger_{\mathbf k\sigma},
\\
\hat a^\dagger_{\mathbf k\sigma}
&=
\sqrt{\frac{ck n_r}{2\hbar}}\,
\hat q^\dagger_{\mathbf k\sigma}
-
i\sqrt{\frac{1}{2\hbar ck n_r}}\,
\hat p_{\mathbf k\sigma}.
\end{aligned}
\end{equation}
With this choice we work within a single reference Fock space with a fixed operator basis throughout the evolution. For notational simplicity, we henceforth suppress the polarization index and the vector notation on \(\mathbf k\).

With this choice, the Hamiltonian for each homogeneous \((k,-k)\) sector takes the quadratic form
\begin{equation}
H_k(t)
=
A(t)(\hat n_k+\hat n_{-k}+1)
+
B(t)
\left(
\hat a_k\hat a_{-k}
+
\hat a_k^\dagger \hat a_{-k}^\dagger
\right),
\label{eq:fullH}
\end{equation}
where \(k\) is the magnitude of the wavevector and \(\hat n_k=\hat a_k^\dagger\hat a_k\). In natural units \(\hbar=c=\epsilon_0=1\) with \(n_r=1\),
\[
A(t)=\frac{k}{2}\bigl[1+\epsilon^{-1}(t)\bigr],
\qquad
B(t)=\frac{k}{2}\bigl[1-\epsilon^{-1}(t)\bigr].
\]

This fixed-basis formulation should be contrasted with instantaneous-basis approaches. In Law's formulation the field is quantized in an instantaneous mode basis, so the associated Fock-space representation changes with time~\cite{law1994effective}. In the fixed-basis formulation used here, the state, field operators, and interaction terms live within a single reference Fock space throughout the evolution. A closely related canonical treatment based on fixed initial modes was also developed in the dynamical Casimir effect context, where the particle representation is specified with respect to the initial modes rather than an instantaneous basis~\cite{kawakubo2011photon}. The fixed-basis language provides a transparent way to track vacuum excitation and light--matter coupling, but the advantage is one of representational convenience rather than principle. 

Because the Hamiltonian remains quadratic, we organize the dynamics at the operator level in the Heisenberg picture. In the present setting, the operator dynamics closes exactly within each independent \((k,-k)\) sector, reducing to a \(2\times2\) linear problem in Nambu space. Introducing
\[
\mathbf\Phi_k=(\hat a_k,\hat a^\dagger_{-k})^\mathsf{T},
\]
the Heisenberg equation \(\partial_t\hat a_k=-i[\hat a_k,H_k(t)]\) can be written as
\begin{equation}
i\partial_t \mathbf\Phi_k=M_k(t)\mathbf\Phi_k,
\qquad
M_k(t)=
\begin{pmatrix}
A(t) & B(t)\\
-B(t) & -A(t)
\end{pmatrix}.
\end{equation}
The Nambu-space evolution operator is
\[
\mathcal U_k(t,0)
=
\mathcal T\exp\!\left[
-i\int_0^t dt'\,M_k(t')
\right],
\]
where \(\mathcal T\) denotes time ordering. Equivalently, \(M_k(t)=\sigma_z H_{{\rm BdG},k}(t)\) with \(\sigma_z={\rm diag}(1,-1)\), where the BdG Hamiltonian is
\begin{equation}
H_{{\rm BdG},k}(t) = \begin{pmatrix}A(t)&B(t)\\ B(t)&A(t)\end{pmatrix}.
\end{equation}
The evolution preserves \(\mathcal U_k^\dagger\sigma_z\mathcal U_k=\sigma_z\), the Nambu-space statement of bosonic commutation-relation preservation. When the modulation is periodic, \(M_k(t+T)=M_k(t)\), and the one-period monodromy \(\mathcal U_k(T,0)\) is the quantum counterpart of the classical monodromy of Sec.~\ref{3-1}. The underlying quadratic Hamiltonian is Hermitian, but \(M_k(t)\) is not: its non-Hermitian form is the direct consequence of the \(\sigma_z\) metric enforcing bosonic commutation relations in Nambu space.

This closed two-mode description is special to the quadratic bosonic sector of a homogeneous, nondispersive medium. Once material dispersion, loss reservoirs, spatial structure, or explicit matter degrees of freedom are included, the dynamics need not remain closed within a single bare \((k,-k)\) block, and a more general treatment is typically required. {In more realistic platforms, this exact closure will generally not survive in its ideal form. Material dispersion can introduce additional microscopic or polaritonic degrees of freedom, and absorption requires reservoir degrees of freedom and the associated noise currents to preserve the canonical commutation relations. Finite and open structures replace bulk momentum sectors by scattering channels, resonant states, or quasinormal modes, and strong material resonances can make the relevant excitations polaritonic rather than purely photonic. The central physical mechanism, temporal positive- and negative-frequency mixing and the associated possibility of amplification, pair correlations, and squeezing, can carry over qualitatively, but photon number, squeezing, and pair creation must then be defined with respect to appropriate input/output, polaritonic, or detector modes. Quantitative predictions for lossy, dispersive, finite, or resonance-dominated implementations generally require a macroscopic-QED, reservoir-based, or input-output formulation rather than the closed two-mode \(SU(1,1)\) model alone~\cite{scheel2008macroscopic,franke2021fermi,ren2021quasinormal}.}

\begin{figure*}[t]
    \centering
    \includegraphics[width=0.9\textwidth]{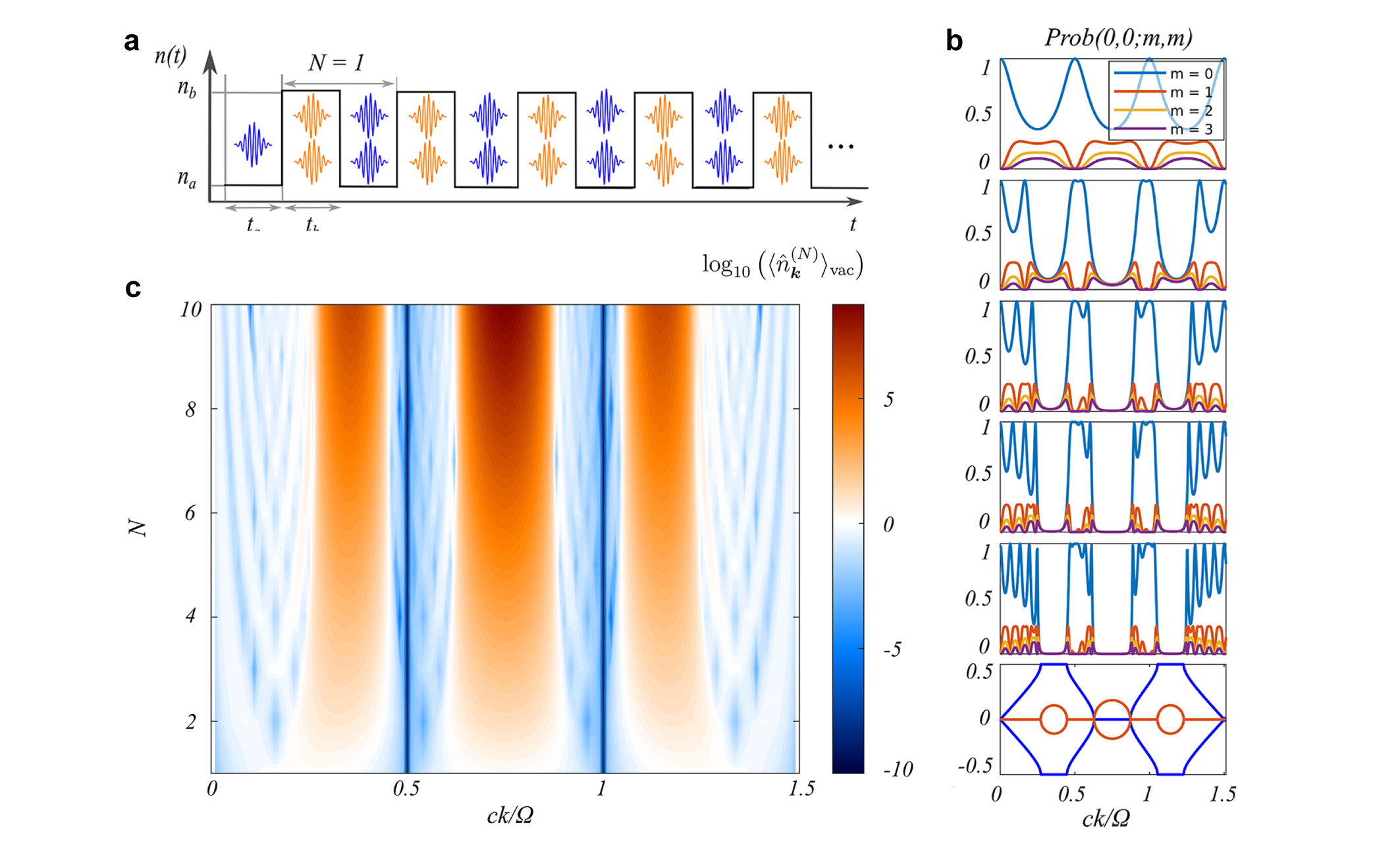}
    \caption[Quantum signatures of vacuum amplification in a PTC]{
    {Theoretical quantum signatures of vacuum amplification in the ideal stepwise PTC model.}
    (a) Schematic of the step-modulated PTC, built from alternating temporal slabs.
    (b) Transition probabilities from the vacuum into the sectors with \(m\) photon pairs (\(m=0,1,2,3\)) for \(N=1,2,3,4,5\) temporal periods (top to bottom), with the corresponding Floquet dispersion shown below.
    (c) Mean photon number extracted from the vacuum, \(\langle \hat n_k^{(N)} \rangle_{\mathrm{vac}}\), as a function of wave number and number of modulation periods, showing exponential growth in the momentum gaps and bounded, oscillatory behavior in the bands.
    {Photon number is defined with respect to the basis convention used in Ref.~\cite{sustaeta2025quantum}.}
   Adapted from Figs.~1, 2, and 3 of J.~E. Sustaeta-Osuna \textit{et al.}, ACS Photonics 2025; \textbf{12}; 1873, with permission from the authors. Copyright \copyright{} 2025 J.~E. Sustaeta-Osuna \textit{et al.}; published by American Chemical Society~\cite{sustaeta2025quantum}.
    }
    \label{fig:sustaeta_ptc}
\end{figure*}

\subsection{Closed BdG dynamics versus open Floquet-QED descriptions}\label{sec:closed_open_floquet_qed}

{The fixed-basis construction above gives a closed, unitary quadratic
description of the field degrees of freedom in a homogeneous PTC with
externally prescribed modulation and no material reservoir. It is a
useful minimal framework for vacuum pair creation, two-mode squeezing,
and momentum-resolved \(SU(1,1)\) dynamics in nondissipative models, but
on its own it does not determine irreversible emitter transition rates,
output spectra, or detector noise in realistic platforms. In periodically
driven open quantum systems, the resulting Markovian generator depends
on whether the bath correlations and secular decomposition are organized
with respect to the undriven Bohr spectrum or the driven Floquet
quasienergy spectrum. Early Floquet--Markov analyses of parametrically
driven dissipative oscillators already emphasized this distinction, and
modern open-Floquet reviews distinguish Floquet--Redfield,
weak-coupling Lindblad, singular-coupling, and phenomenological
approaches~\cite{kohler1997floquetmarkov,mori2022floquetopen}. Which
of these is justified is model dependent, set by the bath spectrum,
system--bath coupling strength, secular approximation, and memory time.
Throughout the closed PTC discussion, exceptional points refer to
coalescences of eigenvectors of the classical Maxwell--Floquet or
bosonic-BdG dynamical matrix.}

This motivates a more careful separation of the descriptions used below. Classical Maxwell--Floquet theory gives band structures and momentum gaps, and it can be formulated in terms of linear-response Green functions. When these response functions are projected into LDOS or kDOS-like quantities in gain-like regimes, the resulting spectral weights can be signed. Closed quantum BdG theory quantizes the ideal quadratic field dynamics and predicts vacuum-seeded pair creation and squeezing in specified modes. Reservoir-consistent Floquet-QED or macroscopic-QED-type descriptions provide the field--matter and reservoir structure, namely Green tensors, noise currents, and properly ordered field correlation functions, from which physical upward and downward transition rates are obtained~\cite{scheel2008macroscopic,franke2021fermi}. For time-varying dispersive media, the relevant macroscopic QED is not the stationary theory with the naive replacement of constitutive parameters, but a reservoir-consistent extension in which the material degrees of freedom and nonequilibrium noise currents are specified explicitly~\cite{horsley2025macroscopic}. An explicit output-channel, scattering, input-output, or equivalent correlation-function description may then be needed to connect intracavity, bulk, or material dynamics to measured output spectra and correlations.

\subsection{Dynamical regimes and quantum observables in PTCs}

{The quantum dynamics of a homogeneous PTC combines two ingredients introduced above: two-mode Bogoliubov mixing generated by temporal modulation, and the Floquet band structure generated by temporal periodicity. In this subsection, we translate this band/gap structure into quantum observables defined with respect to a specified mode convention and reference vacuum. The band, momentum-gap, and gap-edge regimes organize photon number, pair-creation probabilities, squeezing, and anomalous correlations in qualitatively distinct ways~\cite{sustaeta2025quantum,bae2025quantum}.}

In the ideal lossless band regime, where the relevant Floquet quasifrequency is real, the closed two-mode evolution is stable. {Starting from a chosen initial vacuum, pair-sector probabilities and the mean photon number remain bounded and oscillatory~\cite{sustaeta2025quantum,bae2025quantum}.} Within this regime, stroboscopic transparency points can occur, at which transition probabilities into nonzero pair sectors vanish~\cite{sustaeta2025quantum}, and the oscillatory dynamics becomes increasingly pronounced near the gap edge~\cite{bae2025quantum}. {In the momentum gap, where a Floquet quasifrequency branch becomes complex, the vacuum-seeded mean occupation grows exponentially within the two-mode \((k,-k)\) sector, and the pair-number distribution is driven toward progressively higher occupation sectors.} The suppression of low-occupation-number pair states in this regime should not be read as weaker amplification; it reflects a redistribution of probability weight into higher-occupation-number states~\cite{sustaeta2025quantum}. Figure~\ref{fig:sustaeta_ptc} illustrates this distinction between band and momentum-gap dynamics.

{The same band/gap distinction can also be read, within the Floquet-photonic synthetic-space formulation of Ref.~\cite{bae2025quantum}, from a microscopic perspective. In that unbounded model, the band regime corresponds to a localized regime with finite localization length, whereas the momentum-gap regime corresponds to a delocalized regime characterized by Wigner--Dyson level statistics and wave-packet acceleration associated with exponential photonic-energy growth. The gap edges mark the transition between bounded oscillatory dynamics and delocalized amplification dynamics, with the localization length diverging at the critical momenta.}

These observable distinctions are transparent in a fixed operator basis. Once the reference vacuum is fixed and the time dependence is encoded in the Bogoliubov coefficients, or equivalently in the Nambu propagator, one extracts within a single representation the photon number
\[
N_k(t)
=
\langle 0_{\rm ref}|\hat a_k^\dagger(t)\hat a_k(t)|0_{\rm ref}\rangle,
\]
the degree of two-mode squeezing, and the pair-creation probabilities. Within an effective Floquet--BdG description of PTCs, a recent work further argues that the Petermann factor of the corresponding dynamical matrix can be related to Bogoliubov mixing, vacuum quasiparticle population, and the scale of ideal squeezing dynamics and vacuum-fluctuation noise; this remains a theoretical proposal rather than an experimentally established result~\cite{kim2026petermann}. {In a spatially homogeneous bulk PTC, the plane-wave \((k,-k)\) basis is natural because this homogeneity and momentum conservation enforce a counter-propagating two-mode structure from the outset. The fixed plane-wave basis is therefore a particularly transparent representation for relating particle creation from a specified initial vacuum to the classical transfer-matrix description, rather than a basis-independent definition of photon number.} The same closed \((k,-k)\) dynamics can, of course, be rewritten in other equivalent bases. A symmetric/antisymmetric superposition of the \(\pm k\) traveling-wave modes recasts the same two-mode squeezing transformation into an equivalent product of single-mode squeezing transformations, a change of representation rather than of the underlying physical content~\cite{sustaeta2025quantum}.

This equivalence of bases within the homogeneous bulk problem should not be confused with a cavity-DCE reduction. In a cavity, spatial boundary conditions change the mode structure itself, replacing the continuum of propagating $\pm k$ sectors by discrete standing-wave normal modes. The appropriate DCE language is therefore organized around discrete mode-selective parametric resonances rather than bulk momentum gaps. This distinction matters once light--matter coupling is introduced: related band/gap structures can influence emitter observables and local spectral responses, including spontaneous-emission decay, modulation-assisted excitation channels, and Floquet-LDOS-like signatures, but their interpretation as physical transition rates, spectra, or detector-level signals requires the emitter, reservoir, and output-channel assumptions discussed later in Sec.~V.

\section{Comparative frameworks for vacuum amplification in quantum PTCs}\label{sec:DCE_QED}

{In the preceding sections, Bogoliubov mixing was introduced at an idealized temporal boundary and then promoted, by temporal periodicity, to a momentum-resolved Floquet problem for homogeneous PTCs. This section does not rederive those results but places them in a broader interpretive framework. We use the dynamical Casimir effect (DCE), parametric-amplifier language, and spacetime-modulated QED as comparative reference points for vacuum amplification, emphasizing that ideal homogeneous PTCs organize the same quadratic mixing in a distinct momentum-resolved bulk Floquet setting, rather than through the discrete resonances or boundary motion characteristic of standard cavity-DCE settings.}

\subsection{Dynamical Casimir and parametric-amplification perspectives}

\begin{figure}[t]
    \centering
    \includegraphics[width=\linewidth]{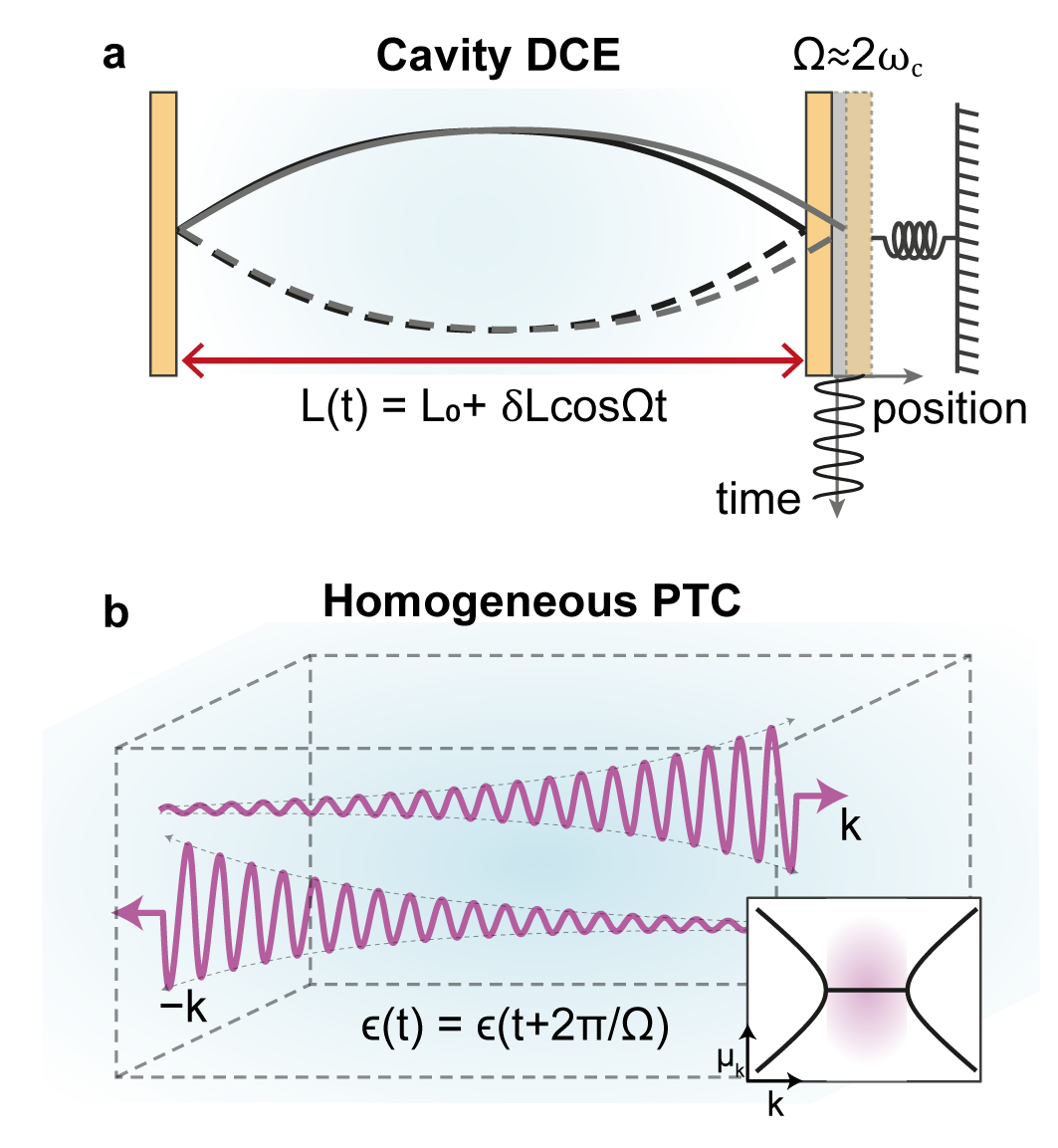}
    \caption{
{Schematic comparison between cavity dynamical-Casimir amplification and homogeneous quantum PTCs.}
{(a) In a resonant cavity-DCE picture, modulation of the cavity length or boundary condition can parametrically amplify vacuum fluctuations in selected discrete cavity modes, typically near a resonance condition such as \(\Omega \approx 2\omega_c\).}
{(b) In an ideal homogeneous PTC, temporal periodicity \(\epsilon(t)=\epsilon(t+2\pi/\Omega)\) couples counter-propagating modes in conserved momentum sectors.}
{The inset is schematic and indicates that, when a momentum gap opens, amplification is organized by a finite region in quasifrequency--momentum space rather than by an isolated cavity resonance.}
}
    \label{fig:dce_ptc_compare}
\end{figure}

{The DCE provides a useful comparative language for photon generation from the vacuum in nonstationary electromagnetic environments, but we use it here as a structural analogy rather than a direct equivalence to PTCs.} Beginning with Moore's variable-length cavity and the moving-mirror analyses of Fulling and Davies, the early moving-boundary formulation established that nonadiabatic changes of electromagnetic boundary conditions can excite the zero-point field and generate photons~\cite{moore1970quantum,fulling1976radiation,davies1977radiation}. This work was later recast in terms of effective quadratic Hamiltonians, making explicit its connection to parametric amplification, photon-pair creation, squeezing, and parametric resonance~\cite{law1994effective,dodonov1996generation,nation2012colloquium,dodonov2020fifty}. {For this comparison, the DCE can be viewed as a family of vacuum-amplification processes in which nonadiabatic modulation of an electromagnetic boundary, cavity, or effective circuit parameter produces Bogoliubov mixing of bosonic modes~\cite{dodonov2020fifty,johansson2009dynamical}.} This picture has been realized experimentally in superconducting-circuit platforms, where rapidly modulated boundary conditions and Josephson metamaterials generated correlated microwave photons and two-mode squeezing~\cite{wilson2011observation,lahteenmaki2013dynamical}. {These experiments serve as quantum benchmarks. They do not realize homogeneous PTCs, but they demonstrate that rapid temporal modulation of an electromagnetic boundary or effective cavity length can generate, directly from vacuum fluctuations, measurable photons and pair correlations and, in related Josephson-metamaterial settings, two-mode squeezed microwave radiation.}

\begin{table*}[t]
\caption{
Comparison of related but nonequivalent frameworks for quantum vacuum amplification.
{In the idealized quantum models listed here, the common algebraic ingredient is Bogoliubov-type annihilation--creation mixing, but the systems differ in their natural degrees of freedom, organizing constraints, and how amplification or instability is selected.}
{The homogeneous quantum PTC is distinguished, when a momentum gap opens, by a momentum-resolved Floquet spectrum with a finite instability window.}
For orientation, the table also includes the single temporal boundary as the elementary Bogoliubov benchmark. Representative references are given in the first column. {The entries summarize schematic, idealized, or reduced descriptions.}
}
\label{tab:framework_comparison}
\small
\renewcommand{\arraystretch}{1.3}
\setlength{\tabcolsep}{3pt}
\begin{tabular*}{\textwidth}{@{\extracolsep{\fill}}lllll@{}}
\toprule
\parbox[t]{2.30cm}{\raggedright\textbf{Framework}}
&
\parbox[t]{2.60cm}{\raggedright\textbf{Natural degrees of freedom}}
&
\parbox[t]{2.70cm}{\raggedright\textbf{Key organizing constraint}}
&
\parbox[t]{3.15cm}{\raggedright\textbf{Where the relevant mixing is organized}}
&
\parbox[t]{3.05cm}{\raggedright\textbf{Main quantum signature}}
\\
\midrule
\parbox[t]{2.30cm}{\raggedright\textbf{Single temporal boundary}~\cite{mendoncca2000quantum,mendoncca2005time,PhysRevResearch.7.013120_2025}}
&
\parbox[t]{2.60cm}{\raggedright Traveling-wave Nambu \((k,-k)\) pair}
&
\parbox[t]{2.70cm}{\raggedright Spatial translation symmetry; conserved wavevector label \(k\)}
&
\parbox[t]{3.15cm}{\raggedright Single temporal boundary; no periodic Floquet instability window}
&
\parbox[t]{3.05cm}{\raggedright Two-mode squeezing, photon-pair creation}
\\ \addlinespace
\midrule
\parbox[t]{2.30cm}{\raggedright\textbf{Cavity/circuit DCE}~\cite{moore1970quantum,law1994effective,dodonov1996generation,nation2012colloquium,wilson2011observation,lahteenmaki2013dynamical}}
&
\parbox[t]{2.60cm}{\raggedright Discrete cavity modes or circuit/waveguide scattering modes}
&
\parbox[t]{2.70cm}{\raggedright Boundary/circuit modulation and resonance matching}
&
\parbox[t]{3.15cm}{\raggedright Resonant cavity/circuit channels or output scattering channels}
&
\parbox[t]{3.05cm}{\raggedright Vacuum photon generation, squeezing, pair correlations}
\\ \addlinespace
\midrule
\parbox[t]{2.30cm}{\raggedright\textbf{DPA-like reduction}~\cite{caves1982quantum,yurke19862}}
&
\parbox[t]{2.60cm}{\raggedright Selected resonant mode in a rotating frame}
&
\parbox[t]{2.70cm}{\raggedright Pump frequency near twice the mode frequency, \(\omega_p \simeq 2\omega_s\)}
&
\parbox[t]{3.15cm}{\raggedright Single-mode gain bandwidth near resonance}
&
\parbox[t]{3.05cm}{\raggedright Single-mode squeezing, phase-sensitive gain}
\\ \addlinespace
\midrule
\parbox[t]{2.30cm}{\raggedright\textbf{NDPA-like reduction}~\cite{caves1985new,schumaker1985new,yurke19862}}
&
\parbox[t]{2.60cm}{\raggedright Selected signal--idler mode pair in a rotating frame}
&
\parbox[t]{2.70cm}{\raggedright Pump-selected frequency matching, \(\omega_p\simeq\omega_s+\omega_i\)}
&
\parbox[t]{3.15cm}{\raggedright Pairwise signal--idler resonance channels}
&
\parbox[t]{3.05cm}{\raggedright Two-mode squeezing, signal--idler correlations}
\\ \addlinespace
\midrule
\parbox[t]{2.30cm}{\raggedright{\textbf{Homogeneous quantum PTC}~\cite{sustaeta2025quantum,bae2025quantum,lyubarov2022amplified}}}
&
\parbox[t]{2.60cm}{\raggedright {Bulk traveling-wave \((k,-k)\) sectors}}
&
\parbox[t]{2.70cm}{\raggedright {Spatial translation symmetry; conserved net momentum \(P_k=\hbar k(n_k-n_{-k})\)}}
&
\parbox[t]{3.15cm}{\raggedright {Momentum-resolved \((k,-k)\) sectors; unstable ones selected by a finite momentum-gap window}}
&
\parbox[t]{3.05cm}{\raggedright {Vacuum-seeded, momentum-resolved many-pair amplification and two-mode squeezing}}
\\
\bottomrule
\end{tabular*}
\end{table*}

{Figure~\ref{fig:dce_ptc_compare} schematically contrasts cavity-DCE and homogeneous-PTC organizations of vacuum amplification.}
Conventional cavity-DCE treatments usually work in an instantaneous cavity-mode basis and often reduce the dynamics to a single resonant mode or a few resonantly coupled modes. {In such a resonant reduction, the effective dynamics can be represented schematically by a single-mode squeezing Hamiltonian,}
\begin{equation}
H_{\text{eff}}(t)
=
\omega(t)\hat a^\dagger \hat a
-i\chi_t\left(\hat a^2-\hat a^{\dagger 2}\right),
\label{eq:DCE_Hamiltonian}
\end{equation}
{where \(\hbar=1\) and the exact prefactors and signs depend on the chosen canonical variables and rotating-frame convention~\cite{law1994effective}.} This Hamiltonian is a reduced cavity-DCE or DPA-like description, not the natural bulk representation of a homogeneous PTC. In a homogeneous PTC, spatial translational symmetry instead organizes the field into independent counter-propagating \((k,-k)\) sectors, each realizing a two-mode Bogoliubov structure, and the instability is selected by a finite momentum-gap window in the Floquet spectrum rather than by an isolated cavity resonance.
In the analogous parametric-amplifier language, cavity DCE and quantum PTCs can be compared with degenerate and nondegenerate parametric amplification, although neither should be identified literally with a conventional DPA or NDPA device. A DPA-like reduction acts on a selected resonant mode driven near twice its frequency, producing phase-sensitive gain and squeezed vacuum~\cite{caves1982quantum,yurke19862}; an NDPA-like reduction acts on a pump-selected signal--idler pair satisfying \(\omega_p\approx\omega_s+\omega_i\), producing two-mode squeezing and signal--idler correlations~\cite{caves1985new,schumaker1985new}. The homogeneous PTC shares this pair-creation algebra, but its natural labels are bulk momenta rather than pump-selected cavity or waveguide channels.
These comparisons also emphasize that mechanical motion is not essential to the common algebraic pair-creation mechanism: within idealized homogeneous time-dependent media, positive- and negative-frequency mixing can produce Bogoliubov mixing and photon-pair creation without a moving mirror~\cite{mendoncca2000quantum,mendoncca2005time,PhysRevResearch.7.013120_2025,prain2017spontaneous}. The temporal-boundary matching used to expose this mechanism is nevertheless model dependent, as discussed in Sec.~\ref{Sec2}, and is not a universal microscopic boundary law~\cite{galiffi2025electrodynamics}. In homogeneous PTCs, temporal periodicity reorganizes the elementary temporal-boundary mixing into a momentum-resolved Floquet band structure, with the momentum gap acting as a band-structured instability window for vacuum amplification~\cite{sustaeta2025quantum}.

\subsection{\texorpdfstring{$SU(1,1)$}{SU(1,1)} perspective and the distinctive structure of homogeneous quantum PTCs}
We now highlight the $\mathfrak{su}(1,1)$ algebraic structure shared by two-mode parametric amplification, resonantly reduced descriptions of cavity DCE, temporal-boundary Bogoliubov transformations, and homogeneous quantum PTCs. Closely related single-mode parametric-amplifier reductions use an analogous single-mode representation of the same algebra.
For a nonzero coupled two-mode sector, such as $(k,\,-k)$ in homogeneous bulk PTC settings, one may define
\begin{equation}
\begin{aligned}
    K_{+}=\hat a_k^\dagger \hat a_{-k}^\dagger,\;
K_{-}=\hat a_k \hat a_{-k},\;\\
K_{0}=\frac{1}{2}\left(\hat a_k^\dagger \hat a_k+\hat a_{-k}^\dagger \hat a_{-k}+1\right),
\end{aligned}
\label{eq:SU11_generators}
\end{equation}
which satisfy the \( \mathfrak{su}(1,1) \) commutation relations,
\begin{equation}
[K_0,K_\pm]=\pm K_\pm,\qquad [K_+,K_-]=-2K_0.
\end{equation}
{The standard two-mode squeezing dynamics can be written in terms of \(K_0\), \(K_+\), and \(K_-\), making \(\mathfrak{su}(1,1)\) a compact algebraic language for parametric amplification~\cite{ban1993decomposition}. In the homogeneous PTC context, this same algebra packages the momentum-resolved two-mode squeezing dynamics~\cite{sustaeta2025quantum}.} {The algebraic closure does not by itself determine whether a given sector is stable or unstable; that is fixed by the physical coefficients, resonance conditions, or Floquet monodromy of the specific system.} {For homogeneous PTCs, a recent work further proposes that the Petermann factor of the dynamical matrix sets the scale of vacuum-fluctuation noise and squeezing dynamics in the same two-mode \(SU(1,1)\) setting; as noted above, this remains a theoretical proposal~\cite{kim2026petermann}.}
{Although these systems share a common $\mathfrak{su}(1,1)$ backbone for pair creation, they are distinguished by the physical constraint that organizes the relevant dynamics.} In the homogeneous bulk case considered here, the medium is cavity-free and translationally invariant, so the natural degrees of freedom are counter-propagating bulk modes $(k,-k)$ rather than discrete cavity eigenmodes. {In the homogeneous Hamiltonian, number-conserving and pair-creation terms coexist, but spatial translation symmetry keeps the total field momentum conserved. For a single \((k,-k)\) block this conservation appears as the invariance of the number imbalance $n_k-n_{-k}$, equivalently of the pair-sector momentum $P_k=\hbar k(n_k-n_{-k})$, so an initial vacuum remains in the balanced ladder $n_k=n_{-k}$ and distinct imbalance sectors do not mix~\cite{bae2025quantum}.} {Unlike DPA/NDPA or cavity-DCE pictures, which emphasize pump-selected or resonance-selected channels, the homogeneous PTC is momentum resolved: when a momentum gap opens, the unstable dynamics is selected over a finite interval of $k$ rather than by a single resonant mode or a pump-selected signal--idler channel~\cite{bae2025quantum,sustaeta2025quantum}.}
{Table~\ref{tab:framework_comparison} summarizes these differences.}
{A related extension of this theme appears in spatiotemporally modulated moving-grating systems, which are adjacent spacetime-QED platforms rather than homogeneous PTCs.}

\subsection{Moving-grating and synthetic-motion viewpoints}

\begin{figure*}[t]
    \centering
    \includegraphics[width=0.95\textwidth]{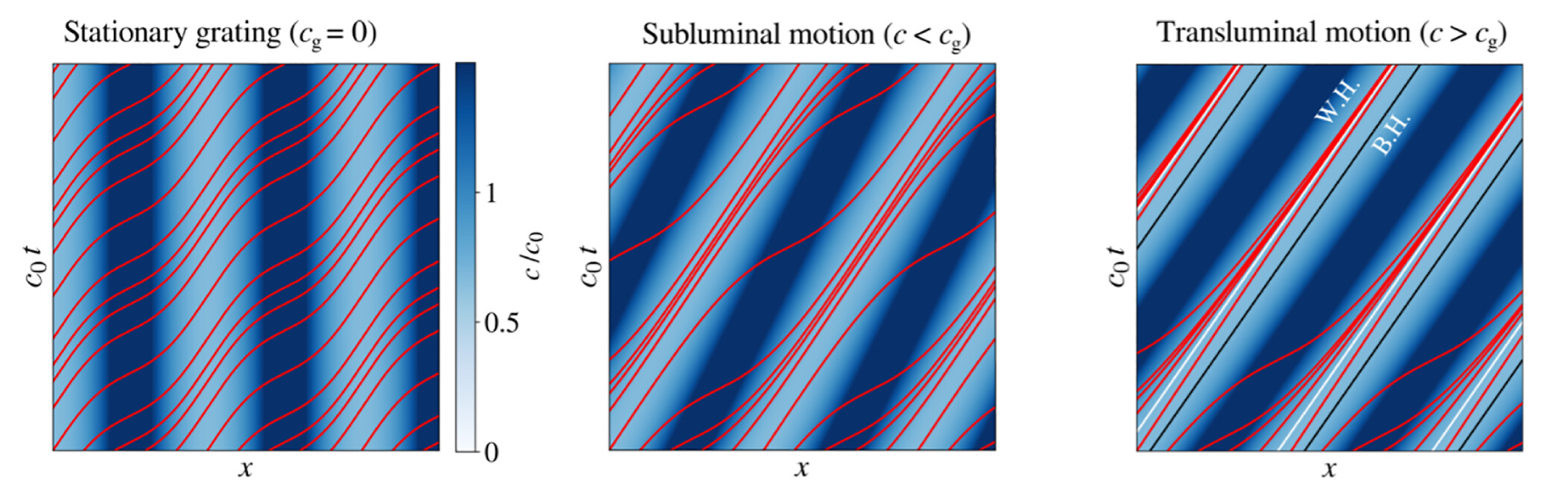}
    \caption{Synthetic-motion picture of a spacetime grating. The stationary, subluminal, and transluminal regimes are compared. In the transluminal case, points at which the grating velocity matches the local wave speed act as optical-horizon analogs in this spacetime-grating geometry, leading to accumulation and depletion of rays. This figure concerns a spatially structured spacetime modulation, not a homogeneous PTC. Reproduced without modification from Ref.~\cite{horsley2023quantum}. Copyright \textcopyright~2023 S.~A.~R. Horsley and J.~B. Pendry. Published by PNAS under the CC BY-NC-ND 4.0 license.}
    \label{fig:moving_grating}
\end{figure*}

{A related extension of the broader vacuum-amplification problem appears in spatially structured, spatiotemporally modulated systems that can be viewed as synthetically moving gratings~\cite{huidobro2019fresnel,prudencio2023replicating}. These systems should be distinguished from the spatially homogeneous PTCs emphasized above.} Figure~\ref{fig:moving_grating} illustrates the basic ray-optics picture.
{In such models, a traveling refractive-index profile of the form $n(x-c_g t)$ represents synthetic motion of the grating pattern rather than literal material motion. Under suitable constitutive assumptions, such spacetime modulations can mimic aspects of moving-media electrodynamics and support Fresnel-drag-like effects, transluminal mode conversion, or effective optical-horizon analogs.} In the transluminal regime, where the grating speed matches the local wave speed at isolated points, accumulation and depletion points emerge that can be interpreted as analog white- and black-hole horizons for light~\cite{horsley2023quantum,pendry2022photon,leonhardt1999optics,philbin2008fiber}.

{Within their finite time-varying-grating model, Horsley and Pendry predicted vacuum emission with Hawking-like features from positive--negative-frequency mixing, together with stimulated pair emission under external illumination~\cite{horsley2023quantum}.} {More broadly, spacetime-modulated media admit QED descriptions in which positive--negative-frequency mixing plays a central role in photon-pair creation~\cite{pendry2024qed}; related superluminal spacetime-boundary platforms such as relativistic plasma mirrors have likewise been predicted to exhibit time reflection/refraction and quantum-vacuum pair generation~\cite{pan2025superluminal}.} {At the macroscopic QED level, these spacetime-modulated settings inherit the same constraint discussed in Sec.~\ref{sec:closed_open_floquet_qed}: time-varying dispersive media require a reservoir-consistent extension with explicit nonequilibrium noise currents, not a naive replacement of constitutive parameters in stationary MQED~\cite{scheel2008macroscopic,horsley2025macroscopic}.}

{A moving-grating problem is organized by a traveling spacetime profile and by the corresponding spacetime-scattering channels, whereas a homogeneous PTC is organized by spatially uniform temporal modulation, conserved $k$-labels, and momentum-resolved coupling within counter-propagating $(k,-k)$ sectors. The latter structure makes a fixed-basis bulk description particularly natural.} {Still, for the quadratic time-varying electromagnetic models discussed here, both frameworks point to the same broader lesson: vacuum amplification is tied to positive--negative-frequency mixing, which appears in the quantum description as Bogoliubov mixing.} {The relation to DCE and spacetime-grating QED is therefore structural rather than identical: the shared ingredient is positive--negative-frequency (annihilation--creation) mixing, while the mode organization, boundary geometry, spectra, and experimentally measured output channels differ from one platform to another~\cite{pendry2024qed,horsley2023quantum}.}

\section{Spontaneous emission, LDOS, and broader platforms in PTCs}

{This section shifts the focus from ideal homogeneous-bulk vacuum amplification to light--matter interactions and experimentally accessible signatures. The central question is no longer whether a homogeneous PTC amplifies vacuum fluctuations, but how its Floquet band, momentum-gap, and gap-edge structure is reflected in emitter response, local probes, and realistic finite or dispersive platforms. Emission-related observables help here, but they must be read at the appropriate level of description: classical power flow, semiclassical linear response, and positive microscopic upward and downward transition rates are not interchangeable.}

Early work on PTCs and related time-periodic photonic media already suggested that momentum-gap physics could lead to amplified radiation, parametric oscillation, or lasing-like behavior~\cite{lee2021parametric,lyubarov2022amplified}. We treat these works mainly as historical motivation rather than as the primary foundation for the emission- and light–matter frameworks discussed below. A broader precursor also existed outside the strict PTC setting: time-modulated photonic-band-gap environments were shown to reshape the spontaneous emission of a single quantum emitter, so temporal modulation of a structured electromagnetic bath can qualitatively modify radiative decay even before the modern homogeneous-PTC framework was fully established~\cite{calajo2017control}. {Against this background, we organize the section around three complementary but nonequivalent levels of description: a classical emitter-based Floquet treatment of spontaneous-emission decay and modulation-assisted excitation in homogeneous PTCs~\cite{park2025spontaneous}, a quantum-electrodynamical model of atom-coupled PTC dynamics and its synthetic-space interpretation~\cite{bae2025quantum}, and experimentally accessible LDOS-related observables in linear-response platforms~\cite{lee2026analogs}. These levels should not be collapsed into a single notion of ``the'' spontaneous-emission rate without specifying the field quantization, noise model, emitter state, and detection observable.}

{Our aim is therefore not to enumerate emission effects but to clarify how bulk Floquet structure is reorganized in passing from homogeneous-bulk vacuum amplification to emitter-resolved, probe-resolved, and material-specific observables. We first review spontaneous-emission decay, modulation-assisted excitation, and atom--photon dynamics in homogeneous PTC models~\cite{park2025spontaneous,bae2025quantum}. We then discuss LDOS-based interpretations and their limitations: LDOS remains a powerful organizing concept in passive, purely lossy linear-response settings, but gain-like response, non-Hermitian Floquet spectra, and nonequilibrium modulation require care beyond the simplest Purcell-like picture~\cite{carminati2015electromagnetic,drezet2017description,franke2021fermi,ren2021quasinormal,ren2024classical}. Finally, we broaden the discussion to finite, dispersive, and material platforms, including temporal-boundary realizations, resonant-state and quasinormal-mode formulations, time-varying resonators for emitter-radiation control, and phonon- or plasmon-mediated realizations that point beyond idealized homogeneous models.}

\subsection{Spontaneous-emission decay, excitation, and atom-coupled synthetic-space dynamics}

{A direct emitter-based analysis of spontaneous emission in PTCs was given by Park
\textit{et al.}, who combined classical light--matter interaction theory with a
Floquet Green-function analysis and found a pronounced enhancement of the
spontaneous-emission decay channel near the momentum-gap frequency of a
PTC. They also identified a second, genuinely
nonequilibrium channel: an emitter initially in its ground state can be promoted
while a photon is emitted, a process they termed spontaneous-emission excitation.
In that picture, the energy deficit is supplied by the externally imposed temporal
modulation of the electromagnetic environment. The scope of this result is
important. Since the emitter is represented as an oscillating dipole coupled to
Maxwell--Floquet fields, the calculation provides a channel-resolved
classical-response estimate of the weak-coupling Purcell/kDOS structure. It is
not, by itself, a microscopic construction of the ordered quantum noise kernels
that would define positive upward and downward transition rates~\cite{park2025spontaneous}.} This channel-resolved classical-response picture is summarized in Fig.~\ref{fig:park}, where the positive- and negative-kDOS regions are associated, respectively, with ordinary spontaneous-emission decay and modulation-assisted spontaneous-emission excitation.

{This energy balance suggests a useful but limited comparison with Unruh-type
spontaneous excitation: an accelerated detector can be excited from its ground
state while radiation is emitted, with the required energy supplied by the
external agent that maintains the acceleration~\cite{unruh1976notes,zhu2007fulling,calogeracos2016spontaneous}.
The PTC case has a different physical source of nonequilibrium energy. The work
is done by the temporal modulation, not by acceleration through a stationary
vacuum. Thus the analogy should be used only at the level of energy balance; it
does not imply a literal
Unruh temperature, a KMS relation, or a detailed-balance structure.}

A broader microscopic interpretation emerges from the quantum-electrodynamical model of Bae \textit{et al.}, in which the transition between the band and momentum-gap regimes of an ideal homogeneous PTC is interpreted as a localization--delocalization transition in Floquet-photonic synthetic space. In that framework, pair creation and annihilation organize the bare photonic problem into momentum-resolved synthetic wires, while the atom--photon interaction expands the accessible Hilbert space into an array of coupled momentum sectors. The same momentum-gap physics that appears classically as amplification is reinterpreted, at the quantum level, as transport in Floquet-photonic synthetic space, and this picture manifests itself in atom--photon dynamics. The term localization--delocalization transition should be read with the assumptions of the ideal unbounded synthetic-space formulation in mind. In finite-time, lossy, or experimentally truncated settings, the corresponding signatures are more conservatively described through growth of photon-number support, synthetic-space wave-packet spreading, and the loss of coherent few-mode oscillations~\cite{bae2025quantum}.

\begin{figure}[t]
    \centering
    \includegraphics[width=0.9\columnwidth]{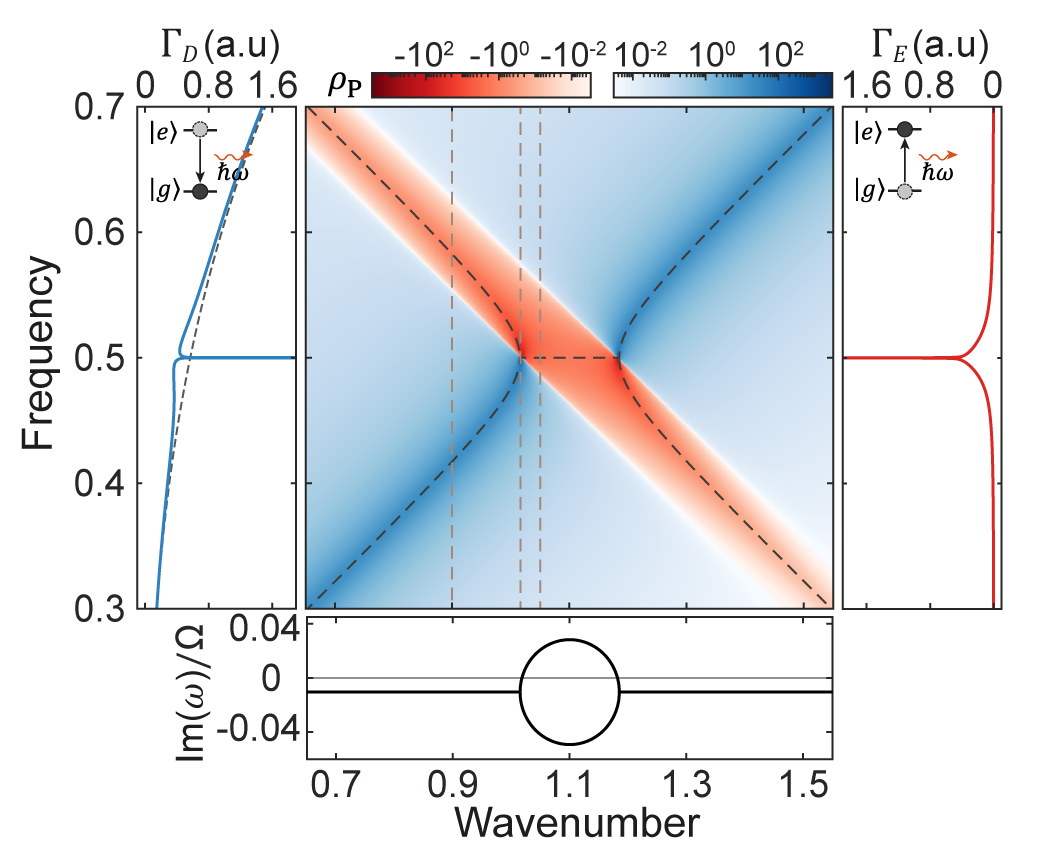}
\caption{Low-loss spontaneous-emission response predicted for a PTC within the classical Floquet light--matter formulation of Ref.~\cite{park2025spontaneous}. The central panel shows momentum-resolved density of states (kDOS), together with the corresponding quasifrequency branches, while the bottom panel shows the imaginary part of the quasifrequency. The left and right panels show the spontaneous-emission decay and excitation channels, $\Gamma_D$ and $\Gamma_E$, respectively. {In that formulation, the negative-kDOS region near the negative-frequency Floquet sideband is associated with the modulation-induced spontaneous-emission-excitation channel; it should not by itself be interpreted as a negative physical decay rate.} Adapted from Fig.~2(e) of J. Park \textit{et al.}, Phys. Rev. Lett. 2025; \textbf{135}; 133801 under the CC BY 4.0 license~\cite{park2025spontaneous}.}
\label{fig:park}
\end{figure}

\begin{figure*}[t]
    \centering
    \includegraphics[width=0.95\textwidth]{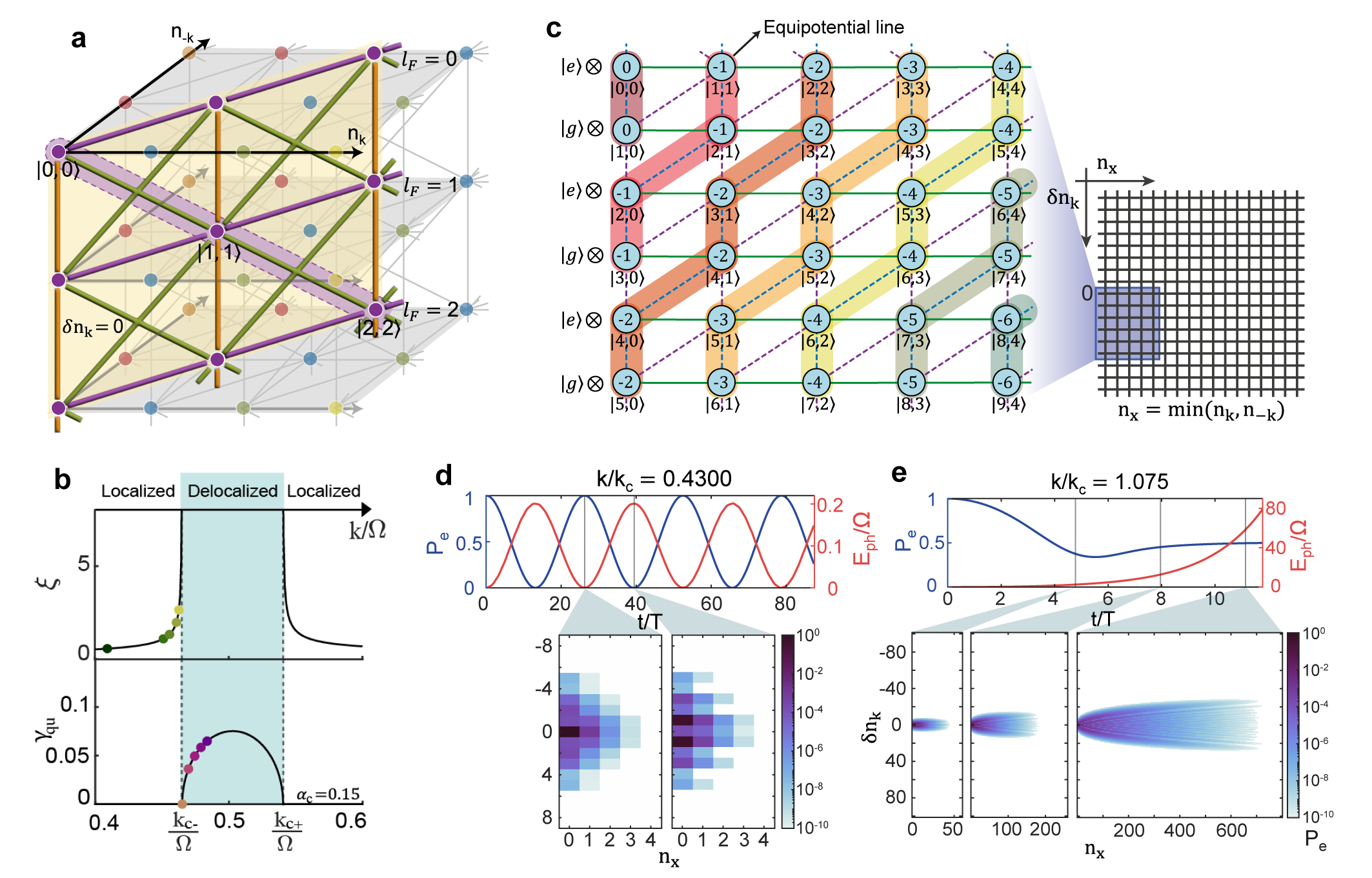}
    \caption{
    Synthetic-space interpretation of light--matter dynamics in a quantum PTC.
    (a) Floquet-photonic synthetic lattice of the bare PTC in the basis $(n_k,n_{-k};l_F)$.
    {(b) Localization length $\xi$ outside the momentum gap and photonic-energy growth rate $\gamma_{\mathrm{qu}}$ inside the gap, as interpreted in the ideal synthetic-space model of Ref.~\cite{bae2025quantum}.}
    (c) Effective atom-coupled Hilbert space, where the photonic synthetic wires are extended by the atomic degree of freedom.
    {(d) For $k<k_{c-}$ in that model, the dynamics remains localized, yielding bounded photonic energy and coherent Rabi-like oscillations.}
    {(e) For $k\in K_{\mathrm{gap}}$, the wave packet delocalizes in synthetic space, the photonic energy grows, and the atomic population relaxes toward an approximately half-and-half mixed state in the idealized calculation.}
    Adapted from Figs.~1, 3, and 4 of J. Bae \textit{et al.}, Nature Communications 2026; \textbf{17}; 858 under the CC BY-NC-ND 4.0 license~\cite{bae2025quantum}.
    }
    \label{fig:bae_qptc}
\end{figure*}

{As summarized in Fig.~\ref{fig:bae_qptc}, the bare quantum PTC in the model of Ref.~\cite{bae2025quantum} is represented in a Floquet-photonic synthetic lattice [Fig.~\ref{fig:bae_qptc}(a)], where the localization length diverges at the critical momenta and the photonic-energy growth rate becomes finite inside the momentum gap [Fig.~\ref{fig:bae_qptc}(b)]. Once a two-level atom is introduced, the relevant Hilbert space becomes an array of coupled momentum sectors [Fig.~\ref{fig:bae_qptc}(c)]. For momenta below the lower critical momentum, the dynamics remains localized, leading to bounded photonic energy and coherent Rabi-like oscillations of the atomic population [Fig.~\ref{fig:bae_qptc}(d)]. By contrast, within the momentum gap the synthetic-space wave packet delocalizes and the photonic energy grows, while the reduced atomic state loses coherent Rabi-like oscillations and approaches an approximately half-and-half mixed state [Fig.~\ref{fig:bae_qptc}(e)]~\cite{bae2025quantum}. In this sense, the atom--PTC dynamics in that idealized setting is naturally understood as a nonequilibrium synthetic-space transport problem, in which pair creation, delocalization, and light--matter coupling jointly govern the loss of coherent Rabi dynamics.}
Related questions of atom--field dynamics in nonstationary cavities have also been explored in the DCE literature \cite{dodonov2011analytical,dodonov2012approximate,dodonov2013continuous}. These studies provide an important conceptual precedent for time-dependent cavity QED.

\begin{figure*}[t]
    \centering
    \includegraphics[width=0.9\textwidth]{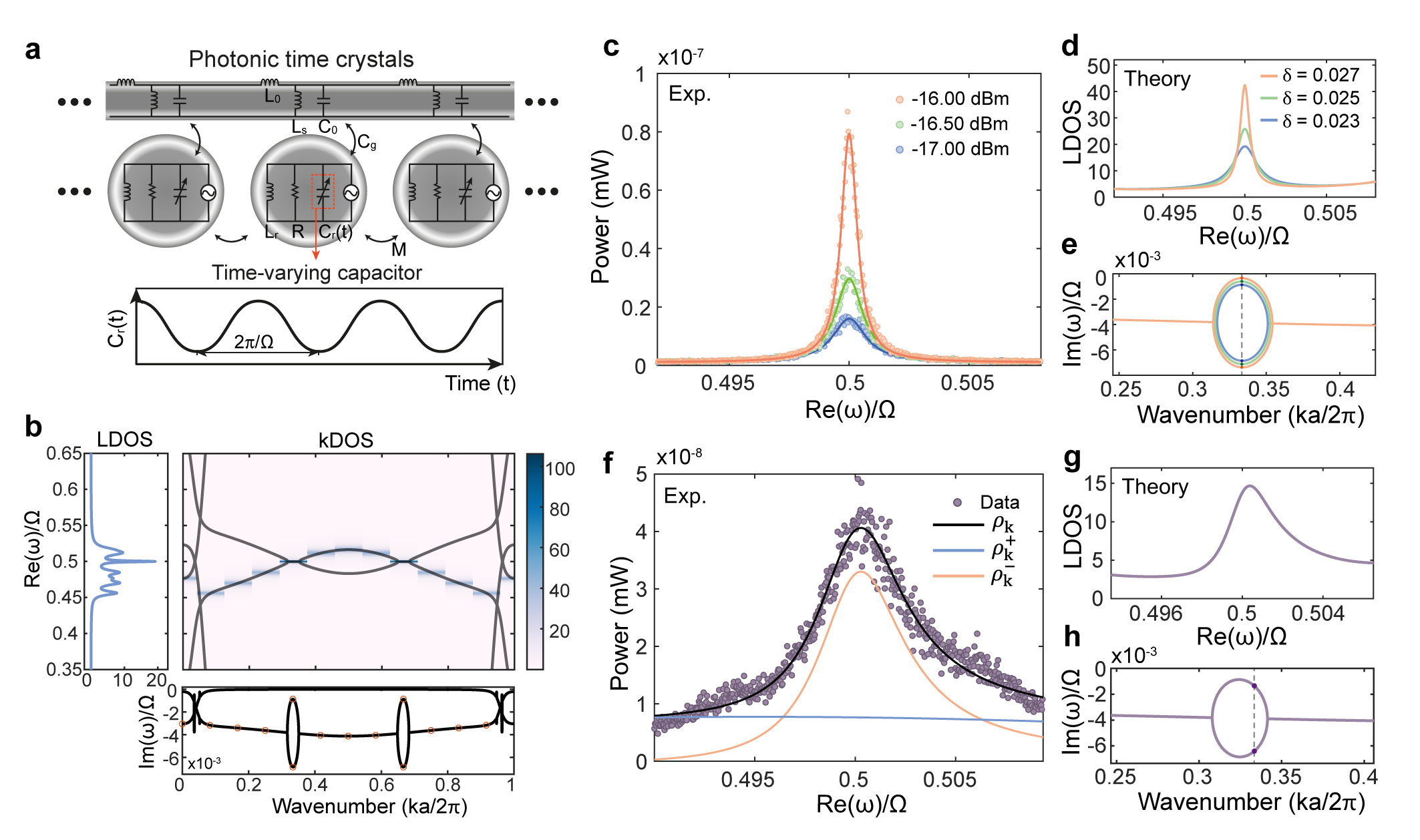}
    \caption{
     {Classical linear-response LDOS and momentum-resolved density of states (kDOS) of a PTC realized with time-periodically modulated LC resonators.}
    (a) Schematic of the experimental platform and sinusoidal capacitor modulation.
    (b) Representative theoretical LDOS (left), kDOS with quasifrequency bands (center), and imaginary parts of the quasifrequencies (bottom), showing the momentum gap modes that generate the LDOS peak near $\mathrm{Re}(\omega)=\Omega/2$.
    (c) Measured subthreshold spectra for gap-center modes at several modulation powers.
    (d),(e) Corresponding theoretical LDOS and imaginary quasifrequency bands for increasing modulation depth, reproducing the peak growth and linewidth narrowing.
    (f) Measured spectrum for a detuned modulation frequency, together with the two-mode fit and the individual contributions of the two in-gap modes.
    (g),(h) Corresponding theoretical LDOS and imaginary quasifrequency bands for the detuned case, reproducing the asymmetric line shape.
    Adapted from Figs.~1 and 3 of K.~Lee \textit{et al.}, Phys.
Rev. Lett. 2026; \textbf{136}; 093802. Copyright (2026) by the American Physical Society~\cite{lee2026analogs}.
    }
    \label{fig:lee_ldos}
\end{figure*}

\subsection{LDOS-based interpretation of emission and its limitations}

{A useful classical-response framework for organizing emission-related phenomena in PTC platforms is the LDOS, which connects the underlying Floquet structure to measurable radiation spectra without, by itself, defining a microscopic quantum transition rate. In this context, Lee \textit{et al.} experimentally probed LDOS-related signatures of a PTC in a classical linear-response setting, using an array of time-periodically modulated LC resonators side-coupled to a bus waveguide [Fig.~\ref{fig:lee_ldos}(a)]~\cite{lee2026analogs}. In the linear, lossy regime, broadband Johnson--Nyquist and residual technical noise excited the system, and the measured subthreshold radiation spectra [Figs.~\ref{fig:lee_ldos}(c) and \ref{fig:lee_ldos}(f)] were consistent with the LDOS-like spectral response predicted by non-Hermitian Floquet theory [Figs.~\ref{fig:lee_ldos}(b), \ref{fig:lee_ldos}(d), and \ref{fig:lee_ldos}(g)]. The momentum gap, arising from the hybridization of positive- and negative-frequency Floquet branches, generated in-gap modes and a sharp LDOS peak near the gap frequency [Fig.~\ref{fig:lee_ldos}(b)]. This peak was nearly symmetric near the gap center [Figs.~\ref{fig:lee_ldos}(c) and \ref{fig:lee_ldos}(d)], but became asymmetric on approaching the gap edge [Figs.~\ref{fig:lee_ldos}(f) and \ref{fig:lee_ldos}(g)], consistent with a decomposition into symmetric and antisymmetric Lorentzian components.}

From a broader theoretical perspective, however, LDOS- or Purcell-like interpretations require some care. {A signed Floquet LDOS or kDOS should not be interpreted directly as a physical transition rate. Physical transition rates must be positive and are obtained from properly ordered quantum field correlation functions, noise kernels, or equivalent master equations. In gain-like or nonequilibrium settings, the signed spectral response must therefore be decomposed into distinct downward and upward channels, rather than treated as a single LDOS-proportional decay rate~\cite{franke2021fermi,ren2021quasinormal}.}
{For PTCs, a more concrete way to formulate this caveat is to examine the momentum-resolved density of states entering the emitter response. A monochromatic dipole with complex amplitude \(\mathbf p\) and frequency \(\omega\) is coupled, through the temporal modulation, to sideband components at \(\omega+n\Omega\), where \(\Omega\) is the modulation frequency.  Accordingly, the relevant LDOS is defined through the frequency-resolved response obtained from the projected Floquet Green function.  In the emitter-based formulation of Ref.~\cite{park2025spontaneous}, this point is made at the momentum-resolved level by defining the dipole-projected kDOS
\begin{equation}
\rho_{\mathbf p}(\mathbf k,\omega)
=
\frac{2\epsilon_0\mu\omega}{\pi}
\operatorname{Im}
\left[
\mathbf n_{\mathbf p}\cdot
G_0(\mathbf k,\omega)
\cdot
\mathbf n_{\mathbf p}
\right],
\label{eq:ptc_kdos}
\end{equation}
where \(\mathbf n_{\mathbf p}=\mathbf p/|\mathbf p|\) is the dipole orientation and \(G_0(\mathbf k,\omega)\) denotes the zeroth Floquet component of the projected dyadic Green's function evaluated at the source frequency \(\omega\).
Within the classical-response prescription of Ref.~\cite{park2025spontaneous}, the sign of the dipole-projected kDOS is used to separate the ordinary decay channel from the modulation-assisted excitation channel. If intrinsic loss is sufficiently large that the relevant kDOS contributions remain positive, the usual weak-coupling power-flow or radiation-reaction argument relates an enhanced LDOS to an enhanced decay-channel response. In a low-loss PTC, however, modulation-induced effective gain can make the kDOS negative near the negative-frequency sideband and inside the momentum-gap region.
In that prescription, the ordinary decay-channel contribution is described by the time-averaged power \(\bar P_D(\omega)\) radiated into the positive-kDOS region \(K_D=\{\mathbf k:\rho_{\mathbf p}(\mathbf k,\omega)>0\}\),
\begin{equation}
 \bar P_D(\omega)
=
\frac{1}{(2\pi)^3}
\frac{\pi\omega^2|\mathbf p|^2}{4\epsilon_0}
\int_{K_D}
\rho_{\mathbf p}(\mathbf k,\omega)\,d^3k.
\label{eq:positive_kdos_decay}
\end{equation}
The decay-channel estimate is then written as \(\Gamma_D(\omega)=\Gamma_0(\omega)\bar P_D(\omega)/P_0(\omega)\), where \(P_0(\omega)\) is the radiated power in the reference homogeneous medium and \(\Gamma_0(\omega)\) is the corresponding reference spontaneous-emission decay rate.
The complementary excitation-channel estimate is obtained from the power-flow contribution associated with the negative-kDOS sector, by integrating the magnitude of the negative kDOS over \(K_E=\{\mathbf k:\rho_{\mathbf p}(\mathbf k,\omega)<0\}\),
\begin{equation}
 \bar P_E(\omega)
=
\frac{1}{(2\pi)^3}
\frac{\pi\omega^2|\mathbf p|^2}{4\epsilon_0}
\int_{K_E}
|\rho_{\mathbf p}(\mathbf k,\omega)|\,d^3k,
\label{eq:negative_kdos_excitation}
\end{equation}
with \(\Gamma_E(\omega)=\Gamma_0(\omega)\bar P_E(\omega)/P_0(\omega)\).
The negative-kDOS region should therefore not be interpreted as a negative decay rate, but as a separate nonequilibrium spontaneous-emission excitation channel in which the temporal modulation supplies the missing energy. Thus, within this classical-response prescription, only the positive-kDOS contribution can be directly associated with ordinary spontaneous-emission decay-rate enhancement.}
{In passive lossy and dispersive media, the LDOS remains a central quantity for weak-coupling light--matter interactions and can be defined rigorously, although its derivation requires a careful macroscopic-QED treatment~\cite{drezet2017description,carminati2015electromagnetic}. A more stringent caveat arises in media containing linear gain, where the usual LDOS-based Fermi's-golden-rule expression for spontaneous-emission can fail and must be replaced by a quantum treatment that separates pump and decay channels or by corrected Purcell-like formulations~\cite{franke2021fermi,ren2021quasinormal,ren2024classical}. This distinction is relevant to PTCs because temporal modulation can generate gain-like response and non-Hermitian Floquet spectra even when the idealized closed microscopic description is Hermitian. Related macroscopic-QED studies of time-modulated media have also highlighted nonequilibrium thermal-emission effects, including nonlocal fluctuation correlations and emission beyond the black-body spectrum, as well as the role of temporal dispersion in regularizing unphysical high-frequency behavior~\cite{vazquez2023incandescent,vertiz2025dispersion}.}

\subsection{Experimental platforms beyond homogeneous PTCs}
{The broader nonequilibrium photonic perspective is further enriched by recent works that move beyond ideal, prescribed homogeneous PTC models toward experimentally accessible temporal boundaries, microwave Floquet platforms, finite structures, and dispersive or material-based time-varying media. Unless stated otherwise, the examples in this subsection should be understood as classical or semiclassical platform demonstrations rather than direct observations of PTC vacuum pair creation.}
{Table~\ref{tab:time_varying_experiments} provides a platform-level classification of the representative examples discussed in this and the following subsection.}
Temporal-boundary realizations are the elementary building blocks of time metamaterials, not PTCs themselves: each demonstrates the single-interface frequency conversion and time reflection that periodic repetition later organizes into a momentum gap. In the terahertz regime, rapidly time-variant metasurfaces have been used to realize linear frequency conversion through a spectrally designed temporal boundary produced by the sudden merging of meta-atoms~\cite{lee2018linear}. A related cavity-enhanced temporal-boundary route was demonstrated using a Fabry--Perot resonator with a temporal-boundary mirror, where an abrupt increase of mirror reflectance and cavity \(Q\) redistributed the spectrum of a coupled THz pulse into the modal frequencies of the post-boundary high-\(Q\) resonator~\cite{lee2022resonance}. At microwave frequencies, switched transmission-line metamaterials have directly observed photonic time reflection and broadband frequency translation at temporal boundaries~\cite{moussa2023observation}, while optically controlled microstrip platforms have observed microwave time reflection and directly probed the phase-conjugation character of the time-reflected wave~\cite{jones2024timereflection}.

\begin{table*}[t]
\caption{
Classification of representative experiments on photonic temporal boundaries,
time-varying photonic media, photonic time crystals, and related time-boundary analogs.
{The table is a platform-level classification: entries in the PTC/Floquet column include classical or semiclassical demonstrations and related Floquet platforms, not direct demonstrations of PTC vacuum pair creation or squeezing.}
}
\label{tab:time_varying_experiments}
\small
\renewcommand{\arraystretch}{1.20}
\setlength{\tabcolsep}{2.5pt}

\newcommand{\TVcellC}[2]{%
\parbox[c]{#1}{\centering\vspace{2pt}#2\par\vspace{2pt}}%
}
\newcommand{\TVcellL}[2]{%
\parbox[c]{#1}{\raggedright\vspace{2pt}#2\par\vspace{2pt}}%
}

\begin{tabular}{@{}cccc@{}}
\hline
\TVcellL{0.20\textwidth}{\raisebox{-0.55\baselineskip}{\textbf{Frequency band}}}
&
\TVcellL{0.34\textwidth}{\raisebox{-0.55\baselineskip}{\textbf{Platform}}}
&
\multicolumn{2}{c@{}}{\textbf{Classification}}
\\
\cline{3-4}
&
&
\TVcellC{0.205\textwidth}{\textbf{Time-interface type}}
&
\TVcellC{0.205\textwidth}{\textbf{PTC/Floquet type}}
\\
\hline

\TVcellL{0.20\textwidth}{Microwave / RF}
&
\TVcellL{0.34\textwidth}{Transmission-line / microstrip platform}
&
\TVcellC{0.205\textwidth}{\cite{moussa2023observation,jones2024timereflection}}
&
\TVcellC{0.205\textwidth}{\cite{reyes2015observation,xiong2025observation,huang2026observationmomentumbandgapphotonic,jones2026broadbandNonresonantTimeCrystal,huang2026microwavevortexbeamlasing}}
\\
\cline{2-4}

\TVcellL{0.20\textwidth}{}
&
\TVcellL{0.34\textwidth}{Coupled-resonator array}
&
\TVcellC{0.205\textwidth}{--}
&
\TVcellC{0.205\textwidth}{\cite{park2022revealing,lee2026analogs}}
\\
\cline{2-4}

\TVcellL{0.20\textwidth}{}
&
\TVcellL{0.34\textwidth}{Time-varying metasurface}
&
\TVcellC{0.205\textwidth}{--}
&
\TVcellC{0.205\textwidth}{\cite{wang2023metasurface}}
\\
\hline

\TVcellL{0.20\textwidth}{THz}
&
\TVcellL{0.34\textwidth}{Time-varying metasurface / photonic cavity / waveguide}
&
\TVcellC{0.205\textwidth}{\cite{lee2018linear,xu2024linearTHzTemporalBoundary,Duan2024,lee2022resonance,Cong2021,PhysRevLett.127.053902}}
&
\TVcellC{0.205\textwidth}{--}
\\
\cline{2-4}

\TVcellL{0.20\textwidth}{}
&
\TVcellL{0.34\textwidth}{Plasmonic / correlated material platform}
&
\TVcellC{0.205\textwidth}{--}
&
\TVcellC{0.205\textwidth}{\cite{guo2026plasmonicmetamaterialtimecrystal,michael2024photonic}}
\\
\hline

\TVcellL{0.20\textwidth}{Optical / near-IR / mid-IR}
&
\TVcellL{0.34\textwidth}{ENZ/TCO thin film}
&
\TVcellC{0.205\textwidth}{\cite{PhysRevLett.120.043902,Zhou2020,tirole2023double,tirole2024second,segal2026timeobservationtimereflectionoptical,jaffray2025spatiospectral,bohn2021spatiotemporalRefraction,lustig2023timeRefractionOptical,PhysRevApplied.18.054067}}
&
\TVcellC{0.205\textwidth}{\cite{galiffi2026coherent}}
\\
\cline{2-4}

\TVcellL{0.20\textwidth}{}
&
\TVcellL{0.34\textwidth}{Electrically tunable metasurface}
&
\TVcellC{0.205\textwidth}{\cite{sisler2024electrically}}
&
\TVcellC{0.205\textwidth}{--}
\\
\hline

\TVcellL{0.20\textwidth}{Related time-boundary analogs}
&
\TVcellL{0.34\textwidth}{Ultracold-atom momentum lattice; fiber-loop time-synthetic lattice}
&
\TVcellC{0.205\textwidth}{\cite{dong2024quantum}}
&
\TVcellC{0.205\textwidth}{\cite{Ren2025}}
\\
\hline

\end{tabular}
\end{table*}

{Optical time-boundary platforms have further extended temporal-boundary physics beyond single-step time refraction. In particular, double-slit time diffraction at optical frequencies was realized by gating a light beam twice in time through the ultrafast modulation of an indium-tin-oxide film, producing interference fringes in the frequency spectrum~\cite{tirole2023double}. Related work demonstrated nonlinear frequency conversion at a time-varying boundary, including second-harmonic generation from an optically pumped ITO film~\cite{tirole2024second}. These optical experiments settle the temporal-boundary and temporal-diffraction ingredients; they do not yet reach the periodic, momentum-gap regime, and they are not by themselves demonstrations of quantum PTC emission.}

A second experimental route is based on microwave, resonator-array, and metasurface platforms that more directly emulate PTC or photonic-Floquet band physics. A historically important precursor was the dynamic-transmission-line experiment of Reyes-Ayona and Halevi, which observed a genuine wave-vector gap in the long-wavelength limit of a temporal photonic crystal~\cite{reyes2015observation}. Building on this microwave-platform direction, a one-dimensional array of time-periodically driven resonators experimentally revealed Bloch--Floquet and non-Bloch band structures of a photonic Floquet medium, establishing a detailed non-Hermitian band-structure description in a microwave realization~\cite{park2022revealing}. Time-varying metasurfaces later extended the PTC concept to two-dimensional artificial structures and experimentally confirmed exponential wave amplification inside a momentum gap in a microwave metasurface design~\cite{wang2023metasurface}. Dynamically modulated microwave transmission-line metamaterials have further enabled the observation of wave amplification within a \(k\)-gap and temporal topological states in a non-synthetic PTC platform~\cite{xiong2025observation}. In a complementary microwave LC-resonator platform, frequency-resolved LDOS-related spectra have been measured, showing LDOS enhancement near the momentum gap and a transition to narrow-band self-oscillation when modulation-induced gain exceeds intrinsic loss~\cite{lee2026analogs}. These platforms close the classical case: the Floquet band, the momentum gap, and the LDOS-like self-oscillation threshold are now measured and reproduced by non-Hermitian Floquet theory. The experimental quantum frontier, however, remains open, requiring calibrated detection of vacuum-seeded correlations and quantum noise.

\subsection{Dispersive, finite, and material extensions beyond homogeneous PTCs}

On the theoretical side, one major direction has been the development of general frameworks for time-varying dispersive media. Quantum-mechanical linear-response theory has clarified how time-dependent response functions are tied to energy transfer and wave propagation in dispersive time-dependent materials, including time-periodic media relevant to dispersive PTCs~\cite{sloan2024optical}. More broadly, dispersion and causality in time-varying media have been revisited at a general level, emphasizing generalized Kramers--Kronig relations and the constraints that realistic material dispersion places on strong temporal modulation~\cite{solis2021functional,koutserimpas2024time}. Complementary operator and eigenmode approaches have also been developed: dispersive time-varying media can be characterized through eigenpulses and operator-valued scattering coefficients~\cite{horsley2023eigenpulses}, while Lorentzian dispersive time-varying media and dispersive--dissipative time-varying systems can be formulated using temporal transfer matrices or Floquet matrix eigenvalue problems~\cite{feng2024temporal,park2025spontaneous,sun2025formulation}. These developments show that the route from ideal nondispersive PTCs to realistic systems cannot ignore the microscopic origin, causality, and spectral structure of the material response. {At the fully quantum level, dispersive time-varying materials require more than the substitution \(\epsilon(\omega)\to\epsilon(t,\omega)\) in stationary macroscopic QED; a reservoir-consistent extension must specify how the material degrees of freedom are modulated and how the associated nonequilibrium noise currents are generated~\cite{horsley2025macroscopic}.}

{A further direction concerns finite, open, and structured implementations, where the relevant physics is no longer governed solely by a bulk momentum-gap picture. Time-varying dispersive boundaries can support natural resonances and boundary-specific frequency-conversion channels that are not captured by a purely homogeneous bulk model~\cite{rizza2024harnessing}. A recent work based on resonant-state theory has argued that temporal modulation in structured PTCs generates Floquet ladders of resonances and that parametric amplification can arise from resonance conditions tied to finite geometries, rather than from the bulk momentum gap alone~\cite{valero2025resonant}. In a related open electromagnetic-scattering formulation, quasinormal modes of Floquet media slabs provide a reduced-order description of electromagnetic scattering, resonant modal hybridization, and gain--loss engineering in time-periodic structures~\cite{vial2025quasinormal}. Together, these approaches indicate that once spatial finiteness, radiation leakage, and resonant confinement are included, the natural organizing objects are resonant states, quasinormal modes, and interface resonances rather than bulk \((k,-k)\) sectors alone.}

Optical material platforms provide another route toward experimentally realistic time-varying photonics. In transparent conducting oxides, ultrafast optical pumping can induce large refractive-index changes on femtosecond timescales. Experiments on aluminium zinc oxide have observed spatio-spectral optical fission of an ultrafast pulse traversing a time-varying subwavelength layer, with the transmitted pulse split into spectrally shifted and angularly separated components~\cite{jaffray2025spatiospectral}. In indium tin oxide, optical coherent perfect absorption and amplification have been demonstrated in a time-varying medium by coherently illuminating a periodically modulated subwavelength film with counterpropagating probe beams~\cite{galiffi2026coherent}. Related optical space--time metasurface experiments have also shown that electrical temporal modulation combined with spatial phase gradients can generate and diffract sideband spectra at telecom wavelengths~\cite{sisler2024electrically}. These systems are not all PTCs in the strict homogeneous-bulk sense, but they identify realistic material and device platforms in which temporal modulation, frequency conversion, and gain/loss-like Floquet response can be engineered.

{A conceptually distinct route has emerged in correlated and strongly driven materials. Michael \textit{et al.} reported that optical excitation of Ta$_2$NiSe$_5$ can generate an unconventional squeezed-phonon state whose fluctuations oscillate at twice the phonon frequency and thereby create an effective Floquet medium in the THz range~\cite{michael2024photonic}. In this picture, the key observable is narrow-band amplification of THz reflectivity, reproduced by a theory based on parametric signal--idler mixing in the driven medium. Ab initio and frozen-phonon calculations identify the 4.7~THz IR-active phonon as the dominant mode responsible for the effect and show that its coupling is strongly reduced in the high-temperature orthorhombic phase, indicating that the photonic time-crystalline response in that material platform is sensitive to the underlying order parameter. A complementary material-based extension has been proposed in plasmonic media, where plasmonic time crystals were predicted to amplify both longitudinal and transverse modes and to support wavevector-independent collective resonances that can persist even in the presence of appreciable dissipation~\cite{feinberg2025plasmonic}. These results suggest, at the level of the cited material-specific models and measurements, that PTC-related amplification can arise in regimes dominated by material resonances, plasmonic dispersion, or correlated-material dynamics, rather than only in weakly dispersive dielectric settings.}

{These material-based routes complement several other recent theoretical extensions beyond homogeneous PTC implementations. Time-varying Mie resonators have been proposed as a route for real-time control over the radiation of quantum emitters~\cite{sadafi2025time}, while recent work on a dispersive plasmonic time-crystal slab has shown that temporal modulation can qualitatively restructure dipole radiation, producing near-field gain, suppression of nonradiative losses, and strong far-field control~\cite{sustaeta2026near}. Related work has predicted broadband dipole absorption in dispersive and absorptive PTCs~\cite{allard2026broadband}, as well as polaritonic amplification in temporally modulated Lorentz--Drude media, where hybrid material resonances and temporal Floquet scattering jointly reshape the instability windows and amplification channels~\cite{ozlu2025floquet}. Taken together, these studies chart the route beyond homogeneous PTCs across both theory and experiment. For a quantum theory of PTCs, they identify an important frontier: extending quantum descriptions beyond ideal homogeneous Floquet media to realistic finite, dispersive, plasmonic, and material-driven time-varying photonic systems.}

\section{Conclusion}

{In this focused review, we have organized the emerging quantum theory of PTCs from temporal boundaries to homogeneous Floquet media and, ultimately, to light--matter dynamics and experimentally accessible observables. Within the homogeneous, nondissipative, source-free setting with prescribed modulation emphasized in the early parts of the review, a temporal boundary can act as a Bogoliubov scatterer that mixes positive- and negative-frequency components within a fixed \((k,-k)\) sector. Temporal periodicity then promotes this single-boundary mechanism to a momentum-resolved Floquet or BdG problem, with homogeneous PTCs providing a bulk band-structured realization. From this viewpoint, pair creation, squeezing, and vacuum amplification are closely related manifestations of positive--negative-frequency mixing in closed quadratic models, rather than independent mechanisms or universal predictions for every time-varying material.}

{This perspective also clarifies how quantum PTCs should be compared with more familiar frameworks for nonequilibrium vacuum amplification. The DCE, DPA/NDPA-like reductions, and \(SU(1,1)\) language remain useful because they isolate the shared pair-creation and squeezing algebra. At the same time, a homogeneous quantum PTC is not exhausted by any one of these analogies: in the ideal bulk model, it is organized as a Floquet problem in which amplification is selected, when a momentum gap opens, by a finite interval of wavevectors rather than by an isolated cavity resonance or a pump-selected signal--idler channel. Moving-grating frameworks provide a related but distinct class of spacetime-modulated QED viewpoints: they share positive-/negative-frequency mixing as a quantum mechanism, but organize the mixing through traveling-profile scattering channels rather than through the momentum-gap structure of a homogeneous PTC.}

{We have also emphasized that this bulk picture becomes richer, and experimentally more consequential, once one moves from vacuum amplification alone to emitter-resolved, LDOS-based, and platform-specific observables. Current experiments on temporal boundaries, microwave Floquet media, metasurfaces, and LC-resonator PTC platforms have established important classical and semiclassical ingredients: temporal scattering and frequency conversion~\cite{lee2018linear,lee2022resonance}, non-Hermitian Floquet band structure and momentum-gap physics~\cite{reyes2015observation,park2022revealing}, momentum-gap amplification in microwave or metasurface PTC platforms~\cite{wang2023metasurface,xiong2025observation}, and LDOS-related spectral response together with above-threshold, lasing-like self-oscillation~\cite{lee2026analogs}. These results provide a realistic classical and semiclassical foundation for pursuing quantum PTC physics, but they are distinct from direct observation of quantum vacuum effects. At the time of writing, the genuinely quantum PTC signatures emphasized in this review---vacuum-seeded photon-pair creation, two-mode squeezing, and nonclassical \((k,-k)\) correlations---remain primarily theoretical targets~\cite{sustaeta2025quantum,bae2025quantum}; a recent work further argues, within an effective Floquet-BdG framework, that classical Petermann factors can be related to squeezing and vacuum-fluctuation noise in PTCs~\cite{kim2026petermann}.}

{This distinction suggests a more concrete experimental outlook. Superconducting-circuit DCE experiments provide a useful quantum benchmark, because vacuum-seeded microwave photon generation, two-mode squeezing, and pair correlations have been measured in time-modulated electrical boundary conditions or Josephson metamaterials~\cite{wilson2011observation,lahteenmaki2013dynamical}; while these experiments serve as benchmarks for time-modulated quantum electrodynamics, they do not demonstrate PTC momentum-gap vacuum amplification. In the near term, microwave PTC and transmission-line platforms are natural experimental testbeds because they allow phase-sensitive detection, correlation measurements, and noise spectroscopy. Direct quantum tests of PTC physics would therefore aim at correlated emission in opposite-momentum or signal--idler channels, quadrature squeezing below the calibrated vacuum reference, momentum-resolved noise spectra, and photon-number statistics across the band, gap-edge, and momentum-gap regimes~\cite{sustaeta2025quantum,bae2025quantum}. LDOS-based microwave experiments and self-oscillation thresholds provide useful classical benchmarks, but the next step is to access fluctuations below or near the single-photon level and to distinguish vacuum-seeded spontaneous pair generation from amplified technical, thermal, or pump-induced noise. In optical and THz platforms, temporal-boundary experiments, time-varying metasurfaces, transparent-conducting-oxide films, and resonant metasurfaces may provide scalable routes to large modulation rates, but quantum detection there will require careful separation of pump-induced noise, material loss, and genuine vacuum-seeded emission.}

{On the theoretical side, the main challenge is to turn the ideal homogeneous quantum PTC into a predictive theory for realistic platforms and observables. Building on the finite, open, and dispersive frameworks described above, one important direction is to extend the fixed-basis quantum formulation to structures where resonant states, quasinormal modes, reservoir dynamics, and noise currents are essential. A second is to connect bulk quantum theory to experimentally measured quantities such as LDOS spectra, output noise, squeezing spectra, and emitter response using open-Floquet, reservoir-consistent time-varying macroscopic-QED, or input-output formulations rather than relying on closed BdG dynamics alone~\cite{kohler1997floquetmarkov,mori2022floquetopen,scheel2008macroscopic,horsley2025macroscopic}. A third is to avoid importing equilibrium Purcell or LDOS formulas into nonequilibrium gain-like Floquet settings without the positive-rate and noise-ordering corrections required in amplifying or mixed loss--gain media~\cite{franke2021fermi,ren2021quasinormal}. Finally, temporal boundary conditions, effective non-Hermitian Floquet descriptions, and synthetic-space models should be connected more directly to microscopic material dynamics in experimentally realistic platforms. This is particularly important for optical and material-based implementations, where dispersion, absorption, pump-induced carriers, phonons, plasmons, and reservoir degrees of freedom may qualitatively reshape the ideal PTC picture.}

{The conceptual structure of quantum PTCs is becoming increasingly clear within idealized models: temporal interfaces generate Bogoliubov mixing, temporal periodicity organizes it into a Floquet momentum-gap problem, and the resulting quadratic field dynamics naturally supports pair creation and squeezing. Once emitter coupling is specified, the same Floquet structure can also feed into modulation-assisted light--matter response. By contrast, the experimental quantum frontier remains open. Establishing unambiguous quantum PTC behavior will require moving beyond classical amplification and band-structure measurements toward calibrated measurements of vacuum-seeded correlations, squeezing, and quantum noise in realistic time-varying platforms. Meeting this challenge would help turn PTCs from an idealized model of nonequilibrium quantum electrodynamics into a practical platform for nonequilibrium quantum photonics.}

\section*{Funding}
This work is supported by the National Research Foundation of Korea (NRF) through the government of Korea (RS-2022-NR070636) and the Samsung Science and Technology Foundation (SSTF-BA240202). K.W.K. acknowledges financial support from the Basic Science Research Program through the NRF funded by the Ministry of Education (no. RS-2025-00521598) and the Korean Government (MSIT) (no. 2020R1A5A1016518).

\section*{Disclosures}
The authors declare that they have no competing interests relevant to this work.

\section*{Data availability}
This article is a review and does not report new experimental or computational data. Accordingly, data sharing is not applicable.
\bibliographystyle{apsrev4-2}
\bibliography{references_main}

\end{document}